%% file: article.tex
\documentclass[aps,prd,twocolumn,superscriptaddress,nofootinbib,floatfix,longbibliography]{revtex4-2}

\pdfoutput=1

\usepackage{amsmath}
\usepackage{dsfont}
\usepackage{graphicx}
\usepackage{float}
\usepackage{bbold} 
\usepackage{xspace} 
\usepackage[caption=false]{subfig} 
\usepackage{grffile}
\usepackage{slashed}
\usepackage{rotating,multirow} 

\usepackage{simplewick} 
\usepackage[prependcaption,textsize=tiny]{todonotes}

\usepackage[hidelinks]{hyperref}
\usepackage[T1]{fontenc}

\graphicspath{{assets/plots}}

\newcommand \spinhalf {spin-$\frac{1}{2}$\xspace}
\newcommand \mpcac {m_{\mathrm{PCAC}}}

\newcommand{\tr}{\mathrm{T}}

\newcommand{\Tr}{\text{Tr}}
\newcommand\pamp{\xi_{\alpha+}}
\newcommand\pamm{\xi_{\alpha-}}
\newcommand{\xib}{\bar{\xi}}

\newcommand\nubar{\overline{\nu}}
\newcommand\Sgroup[2]{\mathrm{#1}(#2)}
\newcommand\su[1]{\ensuremath{\Sgroup{SU}{#1}}}
\newcommand\so[1]{\ensuremath{\Sgroup{SO}{#1}}}
\newcommand\uone{\ensuremath{\Sgroup{U}{1}}}

\newcommand\Nx[1]{N_{\textnormal{#1}}}
\newcommand\Nf{\Nx{f}}

\input{assets/definitions/gammastar_continuumlimit.tex}



\begin{document}


\title{$\su{2}$ gauge theory with one and two adjoint fermions towards the continuum limit}


\author{Andreas Athenodorou}
\email{a.athenodorou@cyi.ac.cy}
\affiliation{Computation-based Science and Technology Research Center, The Cyprus Institute, 20 Kavafi Str., Nicosia 2121, Cyprus}

\author{Ed Bennett}
\email{e.j.bennett@swansea.ac.uk}
\affiliation{Swansea Academy of Advanced Computing, Swansea University, Fabian Way, Swansea SA1 8EN, UK}

\author{Georg Bergner}
\email{georg.bergner@uni-jena.de}
\affiliation{University of Jena, Institute for Theoretical Physics, Max-Wien-Platz 1, D-07743 Jena, Germany}

\author{Pietro Butti}
\email{pbutti@sdu.dk.es}
\affiliation{Departamento de Física Teórica, Facultad de Ciencias and Centro de Astropartículas y Física de Altas Energías (CAPA),
Universidad de Zaragoza, Calle Pedro Cerbuna 12, E-50009, Zaragoza, Spain}
\affiliation{Instituto de Física Teórica UAM-CSIC, Nicolás Cabrera 13–15, Universidad Autónoma de Madrid, Cantoblanco, E-28049 Madrid, Spain}
\affiliation{Quantum Theory Center ($\hbar QTC$) at Department of Mathematics and Computer Science, University of Southern Denmark, 5230 Odense M, Denmark}

\author{Julian Lenz}
\email{j.lenz@hzdr.de}
\affiliation{Swansea Academy of Advanced Computing, Swansea University, Fabian Way, Swansea SA1 8EN, UK}
\affiliation{Helmholtz Zentrum Dresden Rossendorf, Bautzner Landstraße 400, 01328 Dresden, Germany}

\author{Biagio Lucini}
\email{b.lucini@qmul.ac.uk}
\affiliation{Department of Mathematics, Swansea University, Fabian Way, Swansea SA1 8EN, UK}
\affiliation{School of Mathematical Sciences, Queen Mary University of London, Mile End Road,
London, E1 4NS, United Kingdom}


\date{\today}

\begin{abstract}
We provide an extended lattice study of the $\su{2}$ gauge theory coupled to one Dirac fermion flavour ($\Nf =1$) transforming in the adjoint representation as the continuum limit is approached. This investigation is supplemented by results obtained for the $\su{2}$ gauge theory with two Dirac fermion flavours ($\Nf =2$) transforming in the adjoint representation, for which we perform numerical investigations at three values of the lattice spacing. The purpose of our study is to advance the characterisation of the infrared properties of both theories, which previous investigations have concluded to be in the conformal window. For both, we determine the mass spectrum and the anomalous dimension of the fermion condensate using finite-size hyperscaling of the spectrum, mode number analysis of the Dirac operator (for which we improve on our previous proposal) and the ratio of masses of the lightest spin-2 particle over the lightest scalar. All methods provide a consistent picture, with the anomalous dimension of the condensate $\gamma_*$ decreasing significantly as one approaches the continuum limit for the  $\Nf = 1$ theory towards a value consistent with $\gamma_* = \GammaStarContinuumMeanNfOne$, while for $\Nf = 2$ the anomalous dimension converges more rapidly with $\beta$ to a value of $\gamma_*=\GammaStarContinuumMeanNfTwo$. A chiral perturbation theory analysis shows that the infrared behaviour of both theories is incompatible with the breaking of chiral symmetry. 
\end{abstract}


\maketitle


\tableofcontents

\section{Introduction}
\label{sect:introduction}
Large anomalous dimensions have been advocated in beyond the Standard Model strong dynamics to reconcile the proposed fundamental mechanism of electroweak symmetry breaking with experimental observations. A primary example is provided by {\em walking technicolor}, whereby a large anomalous dimension of the chiral condensate would explain the suppression of flavour-changing neutral currents~\cite{Weinberg:1975gm,Susskind:1978ms,Eichten:1979ah,Holdom:1981rm,Holdom:1984sk,Yamawaki:1985zg,Appelquist:1986an}. An additional case in point is partial top compositeness in a theory in which the Standard Model Higgs boson is originated by the global symmetry breaking in a novel strong interaction~\cite{Kaplan:1983fs,Georgi:1984hp,Dugan:1984hq}. In the case of such a Composite Higgs model, a large anomalous dimension of the chimera baryon (a baryon-like composite state containing fermions in the fundamental and in a two-index representation) is needed to enhance the mass of the top quark to the electroweak scale (see., e.g.,~\cite{Barnard:2013zea,Ferretti:2013kya,Ferretti:2016upr,Cacciapaglia:2019bqz}). Large anomalous dimensions are expected to appear in non-Abelian gauge theories near the onset of the conformal window, a phase in which the theory is approximately scale-invariant in the infrared. The opening of the conformal window is determined by the structure and the size of the gauge group, and by the fermion content of the model (specifically, the gauge group representations under which the fermion matter fields transform and the number of fermion flavours in each representation). The onset of the conformal window being a strong coupling problem, analytically only semi-quantitative estimates are possible (see, e.g.,~\cite{Sannino:2003xe,Sannino:2004qp,Dietrich:2005jn,Dietrich:2006cm, Evans:2020ztq,Lee:2020ihn}).

As for other non-perturbative questions in non-Abelian Quantum Field Theories, numerical simulations of the relevant model(s) discretised on a spacetime lattice provide in principle a robust framework to investigate the infrared conformal regime and its onset, enabling us to access crucial quantities such as the value of the anomalous dimension itself. Triggered by the experimental activity that led to the discovery and the characterisation of the Higgs boson~\cite{Aad:2012tfa,Chatrchyan:2012xdj}, the resurgent interest in understanding whether strongly interacting systems beyond the Standard Model can provide a viable fundamental mechanism of electroweak symmetry breaking has spawned a significant number of lattice studies aimed at unveiling the phase diagram of gauge theories for given group families and fermion representations in terms of the size of the gauge group and the number of flavours in each representation (see, e.g.,  Refs.~\cite{DelDebbio:2009fd,Hietanen:2009az,Fodor:2009wk,Bursa:2009we,DelDebbio:2010hu,DelDebbio:2010hx,Fodor:2011tu,Bursa:2011ru,Catterall:2011zf,DeGrand:2011qd,Catterall:2007yx,DelDebbio:2008zf,Hietanen:2008mr,Catterall:2008qk, DelDebbio:2013hha,Hasenfratz:2014rna,Athenodorou:2014eua,Appelquist:2016viq,Hasenfratz:2016dou,Bergner:2016hip, Bergner:2017gzw,Athenodorou:2021wom} and  Refs.~\cite{DeGrand:2012qa,DeGrand:2013uha,DeGrand:2015lna,DeGrand:2015yna,DeGrand:2016pgq,DeGrand:2016htl,Ayyar:2017qdf,Bennett:2017kga,Ayyar:2018zuk,Ayyar:2018ppa,Ayyar:2018glg,Bennett:2019jzz,Bennett:2019cxd,Cossu:2019hse,Bennett:2022yfa,DelDebbio:2022qgu,Bennett:2023wjw,Bennett:2023gbe,Maas:2021gbf,Zierler:2021cfa,Kulkarni:2022bvh} for a necessarily incomplete set of investigations targeting the problem of near-conformality respectively in the context of walking technicolour and top partial compositeness). To this purpose, a number of methods aimed at determining with sufficient accuracy the anomalous dimension of relevant operators or the (scheme-dependent) coupling at which the theory is infrared conformal have been proposed and studied. Among them, popular approaches include hyperscaling, dilaton perturbation theory, scaling of the eigenvalues of the Dirac operator, Schr\"odinger functional, and the gradient flow. Some relevant references to this body of work include~\cite{Lucini:2009an,DelDebbio:2010ze, Appelquist:2017wcg,Appelquist:2019lgk,Patella:2012da,Fodor:2012td, GarciaPerez:2015rda}, with relevant reviews provided for instance in~\cite{Pica:2016ejc,Witzel:2019jbe,GarciaPerez:2020gnf, Appelquist:2022mjb,Rummukainen:2022ekh,Bennett:2023wjw,Lee:2024xhb}. 

While strongly interacting theories beyond the Standard Model provide a clear-cut phenomenological motivation to study near-conformality, the underlying question of the possible phase structure of gauge theories is more fundamental and wider-reaching. The existence of an asymptotically free confined phase and of a Coulomb phase can be inferred directly from the sign of the $\beta$-function as a function of the gauge group size and structure, of the number of flavours and of the representation of the gauge group under which the latter transform, including cases in which fermions transforming in different representations are present. The higher end of the conformal window, before the loss of asymptotic freedom, can be exposed in perturbation theory, giving rise to the well-known Banks-Zaks infrared fixed point~\cite{Caswell:1974gg,Banks:1981nn}. Outside perturbation theory, the conformal window is well understood in the presence of a high degree of supersymmetry~\cite{Seiberg:1994pq}, which imposes strong constraints to the $\beta$-function. In non-supersymmetric quantum field theory, lattice calculations remain the most robust method to investigate the lower end of the conformal window, in which strong coupling plays a crucial role.

In this paper, we investigate two well-known gauge models, both based on gauge group \su{2}, coupled respectively with one and two fermions transforming in the adjoint representation of the gauge group, with the aim of providing further insides on their infrared behaviour and investigate quantitatively the value of the anomalous dimension in the continuum limit. The system with number of fermionic flavours $\Nf = 2$, whose physical relevance relies on the observation that it could provide a minimal template for a realisation of walking technicolour,  has been studied systematically for well over a decade on the lattice, using a wide array of methods. The consensus reached is that the system is infrared conformal with a small anomalous dimension, which excludes it from the class of phenomenologically relevant models. However, a robust determination of the anomalous dimension is still lacking. Tensions in the obtained data are shown for instance in~\cite{Bergner:2016hip}.

The system with $\Nf = 1$  has been considered in the context of semiclassical analysis \cite{Unsal:2007jx} and volume independence, leading to the conjecture of emergent fermion symmetry also in the non-supersymmetric case \cite{Basar:2013sza}. A numerical study of this system was  first provided by some of us in Ref.~\cite{Athenodorou:2014eua},  where it was shown that, despite common lore, the system is not in the expected confining and chiral symmetry breaking phase; moreover, albeit at fixed lattice spacing, this study was the first to determine an order one anomalous dimension for the chiral condensate, which would fulfill one of the requirements of walking technicolour. Subsequent analytic work based on 't Hooft anomaly matching~\cite{Anber:2018tcj,Bi:2018xvr,Cordova:2018acb,Wan:2018djl,Poppitz:2019fnp} showed that the model has an interesting phase structure in which topology plays a key role. An interesting observation made in~\cite{Anber:2018tcj} is the emergence of a light composite fermion in the spectrum. Subsequent numerical studies~\cite{Bi:2019gle,Athenodorou:2021wom} have confirmed that the system does not have the conventional properties of a confining and chiral symmetry breaking theory. However, while more likely, the near-conformal scenario has not proved to provide a striking fit to the data. 
Additionally, the expected light fermion in the spectrum has not been observed, and the anomalous dimension has been found to have a strong dependency on the lattice spacing. 
In contrast to earlier results, a recent study based on an overlap lattice action~\cite{Bergner:2022hoo} favours rather a chiral symmetry breaking scenario. This lattice formulation provides an improved representation of the continuum symmetries, but so far the considered lattices have been rather coarse. A similar result might be deduced from the study of a mixed adjoint-fundamental theory~\cite{Bergner:2020mwl} with Wilson clover fermions, but again the parameter range has been very limited. Finally, the system with $N_f=1$ includes only two Goldstone bosons resulting from the breaking of chiral symmetry. This makes it phenomenologically unviable to account for the standard model electroweak symmetry breaking. However, the system can be extended to a physically viable setup through the ultra minimal walking technicolor (UMWT) model~\cite{Ryttov:2008xe}, which also includes fermions in the fundamental representation.

The current status of understanding of these two models calls for additional simulations on larger lattices, closer to the continuum and to the chiral limit. In alignment with this demand, the purpose of our work is to extend the numerical results of previous investigations in order to provide a more solid understanding of the infrared dynamics of the target models. Therefore, we provide a range of new numerical results for \su{2} gauge theories with $\Nf = 1$ and $\Nf = 2$ adjoint fermion flavours at larger couplings, larger lattices and smaller masses. For continuity with our previous studies, we employ the Wilson action for the gauge part of the models, and we use the standard Wilson discretisation of fermions. This simple discretization allows us achieve very large lattice sizes at small fermion masses and combine a large set of data in our final analysis that includes also those produced in previous investigations. We aim to get to the maximum point that can be achieved with our current approach. At the same time we want to investigate to what extent the same uncertainties that we have observed for $\Nf=1$ also appear in case of the seemingly well investigated $\Nf=2$ theory. The main tools we employ for the characterisation of the model are the behaviour of the chiral condensate, the scaling of the mass spectrum and the scaling of the eigenvalues of the Dirac operator. 

{
The rest of the paper is organised as follows. In Sec.~\ref{sect:model} we provide a quick overview of the continuum theories of interest and their lattice counterpart employed in this work. Section~\ref{sect:results} is dedicated to a discussion of the methodology and the extraction of the main results for the $\Nf = 1$ case, with different physical observables of interest discussed in different subsections. In Sec.~\ref{sect:nf2} we report our numerical results for the $\Nf = 2$ system. Conclusions can be found in Sec.~\ref{sect:con}. Preliminary results of our study have already appeared in~\cite{Bennett:2022bhc}. 
}

\section{Summary of the continuum and lattice theory}
\label{sect:model}

The Lagrangian of the \su{2} gauge theory with $\Nf$ adjoint Dirac flavours in Minkowski space, is defined as:
\begin{align}
\mathcal{L} =& - \frac{1}{2} \mathrm{Tr} \left( G_{\mu \nu}(x) G^{\mu \nu} (x) \right) \nonumber\\
&+ \sum_{\alpha=1}^{\Nf} \overline{\psi}_\alpha (x) \left( i \slashed{D} - m \right) \psi_\alpha(x)\,,\label{eq:actcont}
\end{align}
with $\slashed{D} = \left(\partial_{\mu} + i g A_{\mu}(x) \right)
	\gamma^{\mu}$, $\gamma_{\mu}$ being the Dirac matrices, $A_{\mu}(x) =
	\sum_a T^a A_{\mu}^{a}(x)$ with $a = 1,2,3$, and the $T_a$ the Lie algebra
	generators of $\su{2}$ in the adjoint representation. The field strength tensor is defined as $G_{\mu \nu} = \partial _{\mu} A_{\nu}(x) - \partial
	_{\nu} A_{\mu}(x) + i g [A_{\mu}(x), A_{\nu}(x)]$, with $g$ being the
	gauge coupling of the theory, and the trace is taken
	over the gauge group. Notation and conventions are explained in Ref.~\cite{Athenodorou:2014eua}.

We may re-express each Dirac flavour $\psi_\alpha(x)$ in terms of two Majorana flavours $\xi_{\alpha\pm}(x)$, by making use of the projections
\begin{eqnarray}
\pamp = \frac{\psi_\alpha + C \overline{\psi_\alpha}^\tr}{\sqrt{2}} \ ,
\qquad
\pamm = \frac{\psi_\alpha - C \overline{\psi_\alpha}^\tr}{\sqrt{2}i} \ ,
\end{eqnarray}
such that
\begin{eqnarray}
\psi_\alpha =\frac{1}{\sqrt{2}}( \pamp + i \pamm) \ .
\end{eqnarray}

Eq.~\eqref{eq:actcont} may then be re-expressed as
\begin{align}
\mathcal{L} =& - \hspace{-0.1cm}  \frac{1}{2} \mathrm{Tr} \left( G_{\mu \nu}(x) G^{\mu \nu} (x) \right) \nonumber\\
&+ \frac{1}{2} \hspace{-0.05cm} \sum_{\alpha =1}^{\Nf} \sum_{k\in \{+,-\}} \xib_{\alpha k}(x) \left( i \slashed{D} - m \right) \xi_{\alpha k}(x)\,, \label{eq:actcont2}
\end{align}
exposing the $\su{2\Nf}$ chiral symmetry in the action. The adjoint representation being real, the expected chiral symmetry breaking pattern is 
\begin{align}
\su{2\Nf} \mapsto \so{2 \Nf} \ . 
\end{align}

In this work, we consider the cases $\Nf=1$ and $\Nf=2$.
We use the same computational setup as our previous work~\cite{Athenodorou:2021wom}: we generate dynamical gauge configurations using the standard Wilson plaquette action and the standard Wilson fermion action
\begin{eqnarray}
S=S_{\mathrm{G}} + S_{\mathrm{F}} \,,
\end{eqnarray}
where 
\begin{eqnarray}
  S_{\mathrm{G}} = \beta \sum_{p} {\rm Tr} \left[ 1 - \frac{1}{2}U(p)  \right] \,,
\end{eqnarray}
\begin{eqnarray}
S_{\mathrm{F}} =  \sum_{x,y} {\overline \psi}(x) D(x,y)\psi(y)\,,
\end{eqnarray}
$\Nf$ is the number of flavours, $U(p)$ is the traced parallel transport of the links $U_{\mu}(x)$ around a plaquette $p$, and $D(x,y)$ is the Wilson Dirac operator
\begin{align}\label{eqn:Diracoperator}
  D(x,y) = \delta_{x,y} 
  - \kappa &\left[ \left(1-\gamma_{\mu}  \right) U_{\mu} (x) \delta_{y,x+\mu} \right.\\
    \nonumber 
    &+  \left. \left(1+\gamma_{\mu}  \right) U^{\dagger}_{\mu} (x-\mu) \delta_{y,x-\mu}   \right]\,.
\end{align}
For full details we refer the reader to Ref.~\cite{Athenodorou:2021wom}.
This generation is performed using importance sampling methods with the Rational Hybrid Monte Carlo (RHMC) algorithm~\cite{Kennedy:1998cu} for $\Nf=1$. A change from our previous work is the software used to perform these algorithms: while for heavier ensembles we continue to make use of HiRep~\cite{DelDebbio:2008zf,hirep-github}, for lighter ensembles we use Grid~\cite{Yamaguchi:2022feu,grid-github}, which enables significantly more powerful GPU-based resources to be used. For the two smaller values of $\beta$ for $\Nf=2$, an updated version of HiRep also supporting GPU acceleration was used~\cite{Martins:2024dew}.\footnote{In each case minor modifications were made to the base code referenced above. In the case of HiRep, these are available in the \texttt{su2nf1adj} branch of the fork at \url{https://github.com/edbennett/HiRep}, commit \href{https://github.com/edbennett/HiRep/tree/92bbe4d25feb80ce645bb9225d8bd04b1d67347a}{\texttt{6632758}}. In the case of Grid, they are in the \texttt{su2nf1adj\_dirac13} branch of the fork at \url{https://github.com/edbennett/Grid_lattice}, commit \href{https://github.com/edbennett/Grid_epcc/tree/48ca3a244d3f6ce0caa6510096778f07d80588b3}{\texttt{48ca3a2}}.} Computation of specific observables using these configurations is discussed in the relevant subsections of the next section.

\section{Results for the $\Nf=1$ theory}
\label{sect:results}

\begin{table}
    \caption{\label{tab:ensembles-Nf1-0}
    Ensembles used for the $\Nf=1$ theory,
    $\beta\le2.15$,
    showing the name of the ensemble,
    the coupling $\beta$,
    the bare fermion mass $am$,
    the number of sites in the temporal and spatial directions $N_t$ and $N_s$,
    {the total number of configurations considered $N_{\mathrm{conf.}}$,
    the number of trajectories between each saved configuration $\delta_{\mathrm{conf.}}$,
    and the length of the trajectory in molecular dynamics time units.
    For two ensembles, 
    a dash ``---'' indicates that the trajectory length is no longer known;
    these ensembles were generated for Ref.~\cite{Athenodorou:2014eua}
    and the relevant metadata have been lost.}}
    \scriptsize
    \input{assets/tables/lattice_params_Nf1_part0}
\end{table}

\begin{table}
    \caption{\label{tab:ensembles-Nf1-1}
    Ensembles used for the $\Nf=1$ theory,
    $\beta\ge2.2$,
    showing the name of the ensemble,
    the coupling $\beta$,
    the bare fermion mass $am$,
    the number of sites in the temporal and spatial directions $N_t$ and $N_s$,
    {the total number of configurations considered $N_{\mathrm{conf.}}$,
    the number of trajectories between each saved configuration $\delta_{\mathrm{conf.}}$,
    and the length of the trajectory in molecular dynamics time units.}}
    \scriptsize
    \input{assets/tables/lattice_params_Nf1_part1}
\end{table}

In the following sections we provide an overview of the main results obtained for the \su{2} theory with $\Nf=1$ flavours of adjoint fermion at a wide range of lattice volumes, masses, and coupling constants, as summarised in Tables~\ref{tab:ensembles-Nf1-0} and \ref{tab:ensembles-Nf1-1}. An investigation explaining why we have chosen these parameters to avoid the bulk phase and retrieve the desired continuum physics can be found in our previous investigations~\cite{Athenodorou:2014eua,Athenodorou:2021wom}. In this work the parameter range is significantly larger than in any previous work. In fact, in this work we have pushed the boundary of computational  feasibility with the chosen approach.

Our main goal is to include in particular the dependence on the gauge coupling $\beta$. The large parameter space requires first some general considerations to avoid a breaking of center symmetry indicated by the Polyakov loop or topological freezing. Afterwards we present the results for the scaling of the particle spectrum and the mass anomalous dimension.

\subsection{Expectation value of Polyakov loops and center symmetry}
\label{sec:center_symmetry}
On a finite lattice with compact directions, when the direction $\mu$ is large enough, the model is characterised by a linear realisation of the four center symmetries 
\begin{eqnarray}
\mathbb{Z}_{\mu}(2): U_{\mu}(x) \to \left(1 - 2 \delta_{x_{\mu}, \hat{x}_{\mu}} \right) U_{\mu}(x) \ ,
\end{eqnarray}
which multiplies by $z = -1$ all the links $U_{\mu}(x)$ in the lattice slice identified by the fixed $\mu$-th coordinate $x_{\mu} = \hat{x}_{\mu}$. Here, $\hat{x}_{\mu}$ is a conventionally fixed value. 

The $\mathbb{Z}_{\mu}(2)$ symmetries can be spontaneously broken at small volumes, signaling severe finite-size effects.  
To verify that the finite lattice extent has not caused the centre symmetry of the group to be broken, we compute the expectation value of the Polyakov loop
\begin{equation}
	P_{\mu} = \frac{1}{N}\sum_{x_{\perp}} \mathrm{Tr} \left(\prod_{i=0}^{L_{\mu}-1} U_{\mu}(x_{\perp},x_i)\right)\,.
\end{equation}
A broken centre symmetry would be indicated by two peaks in the histogram of this observable, while the preserved symmetry shows a single peak centered at $\langle P_\mu\rangle=0$. We show a sample of such histograms in Fig.~\ref{fig:polyakov-Nf1}. In all cases we observe this symmetry to be unbroken.

\input{assets/plots/polyakov_Nf1_caption}

\subsection{Topological aspects}
\label{sec:topology}
The formal definition of the topological charge~\cite{Luscher:1981zq,Alexandrou:2017hqw} for a gauge field is given by the four-dimensional Euclidean integral of the topological density across space-time:
\begin{eqnarray}
	Q = \frac{1}{32\pi^2} \int d^4 x \: \epsilon_{\mu\nu\rho\sigma} \Tr\left[G_{\mu\nu}(x)G_{\rho\sigma}(x)\right] \,.
	\label{eq:Q_continuum_def}
	\end{eqnarray}
In practical terms, any appropriate lattice discretization of the topological charge density that results in the correct continuum expression can be employed to assess the lattice equivalence. Using the standard clover representation of the topological charge density, Eq.~\eqref{eq:Q_continuum_def} takes the summation form:
\begin{eqnarray}
	Q_L = \frac{1}{32\pi^2} \sum_{x} \epsilon_{\mu\nu\rho\sigma} \Tr\left[C_{\mu\nu}(x)C_{\rho\sigma}(x)\right]\,,
	\label{eq:Q_fieldteo_def}
\end{eqnarray} 
where $C_{\mu\nu}(x)$ is the standard cloverleaf discretization of the gauge field tensor $G_{\mu\nu}(x)$:
{\small
\begin{eqnarray}
&C_{\mu\nu}(x)&= \frac{\rm Im}{4}\Bigl[
U_{\mu}(x)U_{\nu}(x+ a \hat\mu)U_{\mu}^{\dag}(x+a \hat\nu)U_{\nu}^{\dag}(x) \nonumber \\
&\hspace{-0.65cm} +& \hspace{-0.85cm}
U_{\nu}(x)U^{\dag}_{\mu}(x-a\hat\mu+a\hat\nu)U_{\nu}^{\dag}(x-a \hat\mu)U_{\mu}(x-a \hat\mu) \nonumber \\
&\hspace{-0.65cm} +& \hspace{-0.85cm} U^{\dag}_{\mu}(x-a \hat\mu)U^{\dag}_{\nu}(x-a \hat\mu-a \hat\nu)
    U_{\mu}(x-a \hat\mu-a \hat\nu)U_{\nu}(x-a \hat\nu) \nonumber \\
&\hspace{-0.65cm} +& \hspace{-0.85cm}
U^{\dag}_{\nu}(x-a \hat\nu)U_{\mu}(x-a \hat\nu)U_{\nu}(x+a \hat\mu-a \hat\nu)U^{\dag}_{\mu}(x)
\Bigr]\,.
\end{eqnarray}}

An issue lattice simulations of gauge theories face is the development of significant correlations among topological sectors at reaching the continuum limit. This phenomenon, known as topological freezing, occurs because the probability distribution for the creation of instantons and anti-instantons from the vacuum is suppressed as the continuum limit is approached~\cite{Athenodorou:2021wom}. Consequently, the total topological charge cannot fluctuate adequately, causing the Markov chain to become trapped in specific topological sectors. As a result, the simulation might fail to thoroughly explore every topological sector efficiently and thus be ergodic. In practice, as the continuum limit is approached, pronounced correlations emerge and contribute to an increment of the necessary number of configurations needed to attain statistically significant vacuum expectation values for physical observables.


This issue is a common obstacle in all Markov-chain algorithms where updates are performed locally.  To investigate whether this phenomenon occurs for the values of $\beta$ chosen in our work, we have closely observed the autocorrelation time of the topological charge in our simulation runs.

Monte Carlo assessments of the topological charge $Q$ encounter challenges due to ultraviolet fluctuations that obscure the inherent topological structure. To address this issue, smoothing techniques like gradient flow~\cite{Luscher:2010iy} can be applied to mitigate these fluctuations.

Regarding our simulations for $\Nf=1$, in Fig.~\ref{fig:topcharge_Nf1}, we illustrate the Monte Carlo evolution of the topological charge $Q$ for the ensembles DB4M13, DB5M8, DB6M9, and DB7M10 along with their respective histograms. For DB4M13 DB6M9, and DB7M10 we provide histories for three different threads of configuration production.

These above four sets of configurations correspond to the smallest fermion masses at the four largest values of $\beta$. It is noticeable that none of the ensembles exhibit an exceptionally long autocorrelation time for the topological charge $Q$, indicating that topological freezing is not a concern within the specified parameter range. The simulations explore a substantial number of topological sectors, and the resulting histograms align well with a Gaussian distribution centered at zero. This strongly suggests an effective sampling of topological observables.


In addition to verifying the reliability of the simulations, this study also offers additional insights into the physical properties of the theory. The topological susceptibility $\chi$ defined as
    \begin{eqnarray}
        \chi = \frac{\langle Q^2 \rangle  -  \langle Q \rangle^2 }{V} \,,
    \end{eqnarray}
provides information regarding the potential infrared conformal scenario for the theory. When an infrared (IR) conformal theory is deformed by a fermion mass, it transitions to a confining phase. { For small anomalous dimensions}, the theory should behave similarly to the equivalent quenched theory, with physical bound states becoming highly massive~\cite{DelDebbio:2009fd}. A suitable quantity to verify this scenario is the topological susceptibility, as { in this case} it should align with the results of pure SU(2) Yang–Mills theory~\cite{Bennett:2012ch}. { Conversely, deviations from the quenched value indicate a sizeable anomalous dimension.} 

\begin{figure*}
    \includegraphics[width=\textwidth]{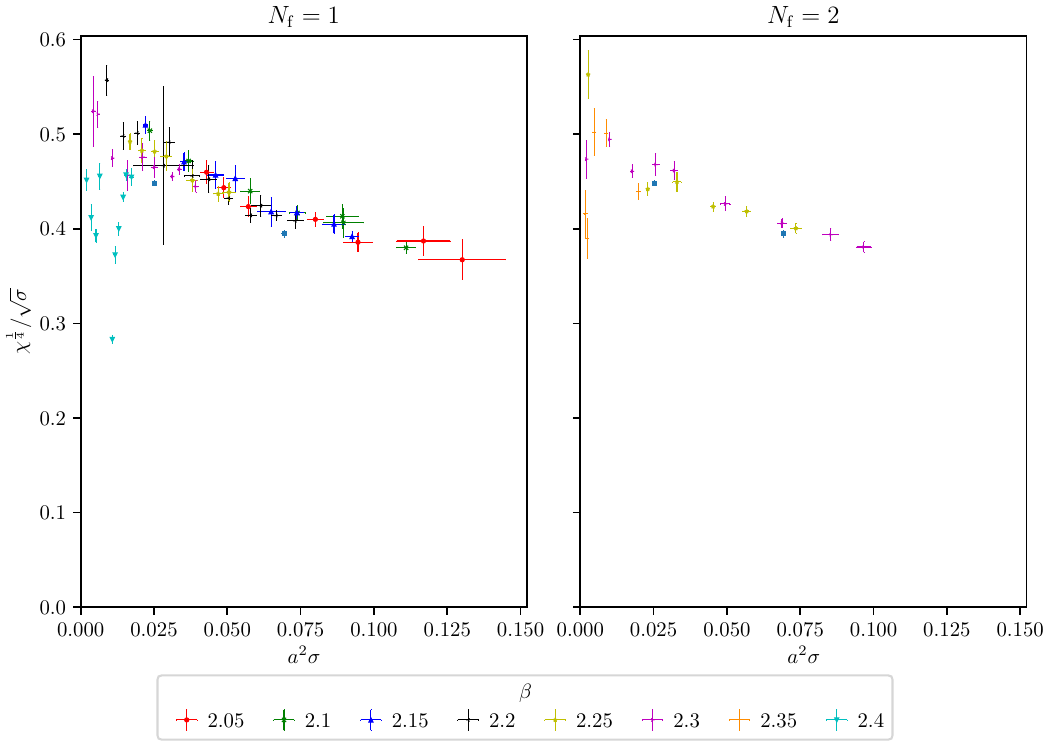}

    \caption{\label{fig:suscept}The topological susceptibility $\chi$ in units of and as a function of the string tension $a^2 \sigma$ for the $\Nf=1$ (left) and $2$ (right) ensembles presented in this work, with $\beta$ values as shown in the legend, compared with data from previous work for the quenched theory~\cite{Bennett:2012ch} (blue squares).}
\end{figure*}

In Fig.~\ref{fig:suscept}, we display the topological susceptibility in units of the string tension, $\chi^{\frac{1}{4}}/\sqrt{\sigma}$, as a function of the string tension in lattice units, $a^2 \sigma$. The results are compared to the corresponding quantity in pure $\su{2}$ Yang–Mills theory for the quenched $\Nf = 0$ ($am \to \infty$) case.
The topological susceptibility for $\Nf=1$ and $\beta \neq 2.4$
{lies close to but systematically above the pure gauge case,
with the same gradient;
this separation may be suggestive of a relatively large anomalous dimension,
or that the theory lies outside the conformal window entirely}.
Clearly,
for $\beta = 2.4$, the topological susceptibility exhibits large negative deviations as we approach $a^2 \sigma \to 0$;
these are understood as underestimated values attributed to critical slowing down,
and are included in the plot for completeness.

In addition, the gradient flow facilitates the establishment of scale parameters, namely, $t_0$ and $w_0$, which can be determined with a high degree of precision using the following prescription: first, set $\mathcal{E}(t)$ as
        \begin{eqnarray}
        \mathcal{E}(t) = t^2 \langle E(t) \rangle \, \ {\rm where} \ E(t) = \frac{1}{4} B^2_{\mu \nu} (t)\,.
    \end{eqnarray}
In this expression $B_{\mu \nu}$ represents the field strength derived from flowing $G_{\mu \nu}$. We designate the scale $t_0$ as the specific value of $t$ at which the following identification occurs:
    \begin{eqnarray}
        \mathcal{E}(t) |_{t=t_0(c)} = \mathcal{E}_0\,. 
    \end{eqnarray}
In a similar manner we define $w_{0}$ in the following way:
     \begin{eqnarray}
        t \frac{\mathrm{d} }{\mathrm{d}t}\mathcal{E}(t) |_{t=w^2_0(c)} = \mathcal{W}_0\,. 
    \end{eqnarray}

In the above definitions, the parameters $\mathcal{E}_0$ and $\mathcal{W}_0$ are selected to ensure that the pertinent condition, either $a \ll \sqrt{8t_0} \ll L$ or $a \ll \sqrt{8w_0} \ll L$, is fulfilled. Choosing a small value of $\mathcal{E}_0$ (or $\mathcal{W}_0$) tends to result in larger lattice artifacts, while larger values typically lead to increased autocorrelations~\cite{BMW:2012hcm}. In our specific scenario, we set $\mathcal{E}_0 = \mathcal{W}_0=0.2$, aligning with the common choice of $\mathcal{E}_0 = \mathcal{W}_0 = 0.3$ in QCD, assuming a scaling with $N$ as described in Ref.~\cite{Ayyar:2017qdf}. $G_{\mu \nu}$ is computed via the clover plaquette $C_{\mu\nu}$.

\input{assets/plots/q_topology_Nf1_caption}

\subsection{Glueball Masses}
\label{sec:glueballs}
Glueball masses can be determined by employing the standard decomposition method on a Euclidean correlator that incorporates an operator referred to as $\phi(t)$~\cite{Athenodorou:2020ani}. This decomposition procedure hinges on expressing these physical states in the framework of the Hamiltonian, designated as $H$, and its corresponding energy eigenstates:
\begin{align}
\langle \phi^\dagger(t=an_t)\phi(0) \rangle & =  \langle \phi^\dagger e^{-Han_t} \phi \rangle \nonumber \\
& = \sum_i |c_i|^2 e^{-aE_in_t} \nonumber \\
& \stackrel{t\to \infty}{=}
|c_0|^2 e^{-aE_0n_t}\,.
\label{extract_mass}
\end{align}
In the above context, $E_0$ signifies the ground state energy. The summation mentioned is constrained to states with non-zero overlaps, satisfying the condition $c_i = \langle {\rm vac} | \phi^\dagger | i \rangle \neq 0$. The quantum characteristics of the operator $\phi$ must align with those of the specific state under examination; i.e., glueball states. Identifying the ground state relies on two crucial factors: the strength of its correlation with the state created by $\phi$ and the rate of exponential decay, as indicated in Eq.~\eqref{extract_mass}. Improving this correlation involves devising operators that effectively capture the essential traits of the state. To enhance the overlap onto the states under investigation, the GEVP technique~\cite{Luscher:1984is,Luscher:1990ck,Berg:1982kp} is employed on a set of operators $\phi_i$ derived from various lattice loops at different blocking levels~\cite{Lucini:2004my,Teper:1987wt}. This process utilizes correlation matrices, denoted as $C_{ij} = \langle \phi_i^{\dagger} (t) \phi_j (0) \rangle$, where $i,j=1,...,N_{\rm op}$, in conjunction with GEVP. Here, $N_{\rm op}$ represents the number of operators used.

\begin{figure}
    \centering
    \includegraphics[width=\columnwidth]{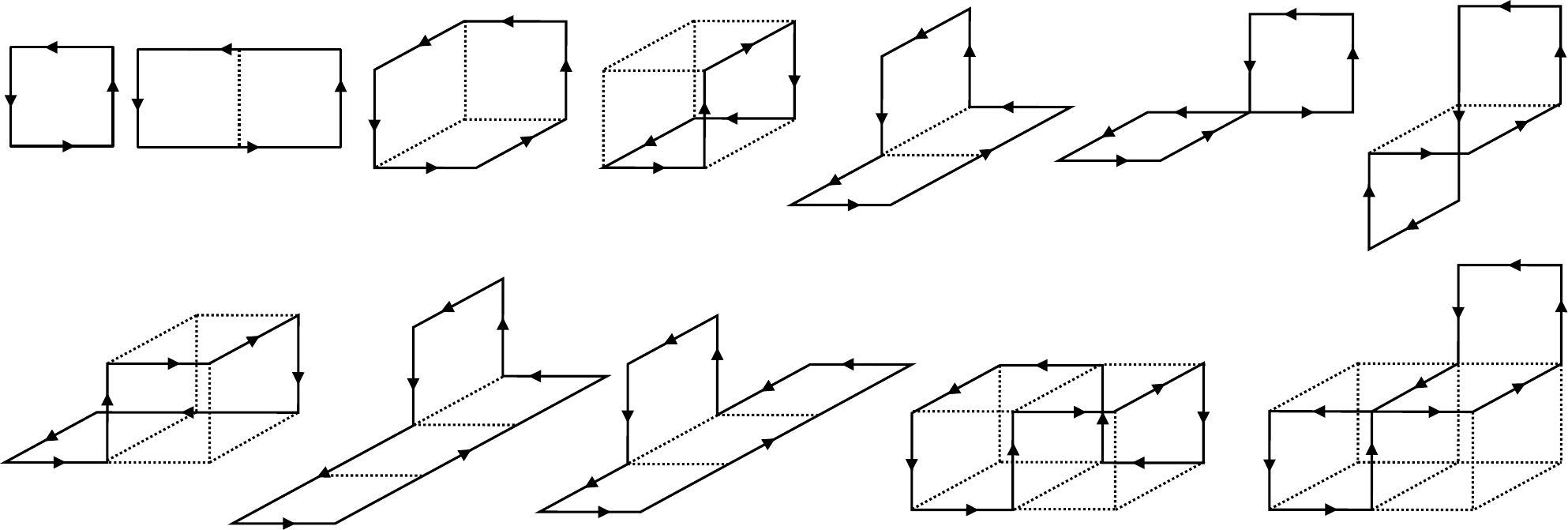}
    \caption{All the different paths used for the creation of the glueball operators}
    \label{fig:glueball_operators}
\end{figure}

To formulate an operator projecting onto a glueball state, we generate an ordered product of $\su{2}$ link matrices along a loop that allows continuous contraction, followed by the calculation of its trace. The real (imaginary) part of this trace corresponds to positive (negative) charge conjugation $C=+$($-$). To ensure the operator possesses zero momentum, we sum over all spatial translations of the loop. Additionally, we consider all possible rotations of the loop, combining them in manners consistent with the irreducible representations ($\mathcal{R}$) of the rotational symmetry group. For the creation of operators with both parities ($P=\pm$), we construct the parity inverse for each loop and then combine them in appropriate linear combinations. In Fig.~\ref{fig:glueball_operators}, we present a selection of the paths utilized in building our basis.

The distinct representations $R$ of the discrete subgroup of cubic rotations within the complete rotation group are labeled as $A_1$, $A_2$, $E$, $T_1$, $T_2$. The $A_1$ representation is a singlet and exhibits complete cubic rotational symmetry, encompassing the $J=0$ state in the continuum limit. Similarly, the $A_2$ representation is also a singlet. The $E$ representation constitutes a doublet, while both $T_1$ and $T_2$ representations are triplets. In the lattice context, the three states corresponding to the triplet of $T_1$ and $T_2$ are degenerate. To address this, their values are averaged, treating them as a single state when estimating glueball masses. The same approach is applied to the $E$ doublets, where their mass estimates are averaged.

The rotational symmetry representations outlined above are rooted in our cubic lattice formulation. As we approach the continuum limit, these states will converge to continuum glueball states falling into representations of continuous rotational symmetry. Consequently, they will form degenerate multiplets comprising $2J + 1$ states, where $J$ denotes the spin of the states. When determining the continuum limit of the low-lying glueball spectrum, it is more informative to assign states to a specific spin $J$ instead of representations of the cubic subgroup. The latter provides a less precise ``resolution," as it maps all spins $J = 1, 2, 3, \dots, \infty$ to just 5 cubic representations. For low values of $J$ ($J=0,1,2$), the distribution of the $2J + 1$ states can be characterized as $A_1 \to J=0$, $T_1 \to J=1$, and $E, T_2 \to J=2$.

\subsection{Torelon masses and the string tension}
\label{sec:string_tension}
Similar to the glueball mass calculation, we can determine the mass of the closed spatial Polyakov loop, also known as a torelon, which wraps around a compactified spatial direction. Torelon states are characterized by quantum numbers such as angular momentum-spin $J$, transverse parity $P_{\perp}$, and longitudinal parity $P_{||}$. The ground state, with $J^{P_{\perp}, P_{||}}$=$0^{++}$, allows for the easiest extraction of the string tension~\cite{Athenodorou:2010cs}. 

The mass of the torelon is determined by computing the ground state energy $m_T(L)$ of a flux-tube with length $L=N_s$, which forms a closed loop by winding once around a spatial compactified torus. We utilize Eq.~(\ref{extract_mass}), where the operator $\phi$ represents the product of $\su{2}$ link matrices along a non-contractible closed path, which wraps around the spatial torus once. We use the most basic form of such an operator, the elementary Polyakov loop. This is the path-ordered product of link matrices in a spatial direction, winding once around the torus along the afforementioned spatial direction. Once again, we utilize smearing and blocking techniques to improve the projection onto physical states. We also sum over translations along the spatial-torus as well as over the transverse directions within the time slice to project onto zero longitudinal and transverse momentum, respectively. As all spatial lattice extents are identical, we average the correlation function over the three spatial directions.

Confining torelons exhibit behavior resembling that of bosonic strings~\cite{Athenodorou:2010cs}. It is expected that the closed flux-tube behaves akin to a confining bosonic string, whose spectrum can be approximated with reasonable accuracy by the Nambu-Goto string model~\cite{GODDARD1973109}. The ground state mass of the closed bosonic string is described by the expression:
\begin{eqnarray}
m_T(l) = \sigma l \sqrt{1 - \frac{2 \pi}{3 \sigma l^2} }.
\label{eq:nambu_goto}
\end{eqnarray}
Here, $\sigma$ represents the string tension and $l=a L$ indicates the length of the torelon. Therefore, we calculate the string tension across the torelon mass spectrum. Results for the string tension are shown in Fig.~\ref{fig:masses_Nf1}.

\subsection{States involving fermions}\label{ssec:states_involving_fermions}

\begin{table}
    \caption{\label{tab:channels}A summary of the nomenclature used for meson symmetry channels in this work, including the interpolating operator used to access each channel, the quantum numbers under the $\uone$ baryon symmetry and parity, and the name we use to refer to the state in the theories presented here. (The vector meson, marked *, is in fact accessible from the connected correlation function, as the disconnected contribution is identically zero.)}

    \begin{tabular}{cccc}
        $\Gamma$ & $\uone^P$ & Name \\
        \hline\hline
        $\gamma_5$, $\gamma_0 \gamma_5$ singlet & $0^-$ & Pseudoscalar meson \\
        1 triplet & $\pm 2^-$ & Pseudoscalar (anti)baryon \\
        1, $\gamma_0$ singlet & $0^+$ & Scalar meson \\
        $\gamma_5$, $\gamma_0 \gamma_5$ triplet & $\pm2^+$ & Scalar (anti)baryon \\
        $\gamma_5 \gamma_k$, $\gamma_0 \gamma_5 \gamma_k$ singlet & $0^+$ & Axial vector meson \\
        $\gamma_k$, $\gamma_0\gamma_k$ singlet* & $0^-$ & Vector meson \\
        $\gamma_5 \gamma_k$ triplet & $\pm2^-$ & Vector (anti)baryon
    \end{tabular}
\end{table}

Mesons and baryons are sourced by operators of the form
\begin{equation}
    O(x) = \overline{\psi}(x) \Gamma \psi(x)\;,
\end{equation}
where $\Gamma$ is a product of Dirac $\gamma$ matrices giving the symmetry of interest.
As discussed above,
this theory has a chiral symmetry breaking pattern
$\su{2\Nf}\rightarrow \so{2\Nf}$,
and we choose to label the states observed in these terms.
The remaining subgroup $\so{2}$ in case of $\Nf=1$ is equivalent to $U(1)$ (refer to as $U(1)$ baryon symmetry).
For convenience,
in Table~\ref{tab:channels} we summarise this convention.

Similarly to the case for glueballs above,
correlation functions of these operators
converge at large time to an exponential decay
with the ground state mass in the exponent
\begin{equation}
    \langle O_X^\dagger (t) O_X(0) \rangle \stackrel{t\rightarrow \infty}{=} f_X\cdot e^{-m_X t}\;,\label{eq:correlator}
\end{equation}
where $m_X$ and $f_X$
are the ground state mass and (unrenormalised) decay constant of
the state sourced by $O_X(x)$.
In a finite volume with antiperiodic boundary conditions in the temporal direction,
this then becomes
\begin{equation}
    \langle O_X^\dagger (t) O_X(0) \rangle \stackrel{t\rightarrow \infty}{=} f_X \left[e^{-m_X t}+e^{-m_X (T-t)}\right]\;,\label{eq:symmetric-correlator}
\end{equation}
Inverting Eq.~\eqref{eq:symmetric-correlator} defines an effective mass $m_\mathrm{eff}$,
and by finding where $m_{\mathrm{eff}}$ has a plateau at large time,
we may identify the region that is free from the effect of excited states.
We then fit the correlator to Eq.~\eqref{eq:symmetric-correlator}
to obtain the mass and decay constant of the state.

Additionally,
as the bare Wilson fermion mass acquires an additive renormalisation,
we define a partially renormalised fermion mass---the PCAC mass---via the axial Ward identity~\cite{DelDebbio:2007wk},
that is expected to be zero in the chiral limit.

In this work we continue to use HiRep to implement 
the computation of correlation functions.
The fits are performed using the \texttt{pyerrors}~\cite{Joswig:2022qfe} library,
which computes uncertainties using the $\Gamma$ method~\cite{Wolff:2003sm},
automatically accounting for autocorrelations in the data.

One particular difference to theories in the fundamental representation is 
the appearance of bound states formed out of gluons combined with
an adjoint fermion field.
One particular of these states is a \spinhalf field 
(also called hybrid fermion,
and denoted as $\breve{g}$ in plots),
which corresponds to the so-called gluino-glue field in $\mathcal{N}=1$ supersymmetric Yang-Mills theory.
The corresponding operator is 
\begin{equation}
    O(x) = \sum_{\mu\nu}\sigma_{\mu\nu}\Tr \left[ F^{\mu\nu}(x)\lambda(x) \right],
\end{equation}
where $\sigma_{\mu \nu}=\frac{1}{2} \left[ \gamma_{\mu}, \gamma_{\nu}  \right]$, the field strength is represented by a clover plaquette and smearing is applied to improve the signal.
Further details of the techniques can be found in corresponding publications of supersymmetric Yang-Mills theory \cite{Bergner:2012rv}.
This correlation function is
computed separately from the spectrum of mesons and baryons
using the FlexLatSim code~\cite{Ali:2018dnd},
but analysed with the same workflow using \texttt{pyerrors}.

\begin{figure*}
    \includegraphics[width=\textwidth]{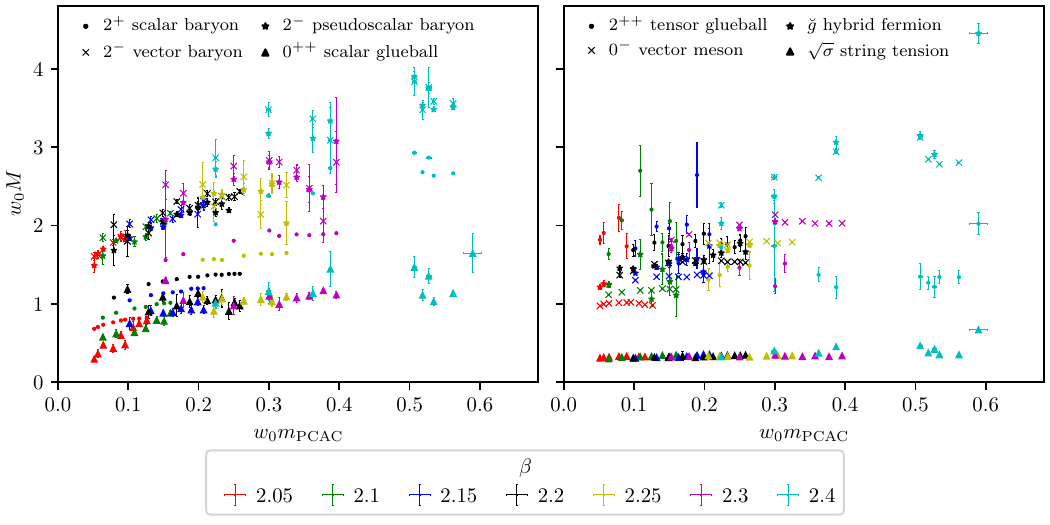}

    \caption{\label{fig:masses_Nf1}Spectrum of particle masses in the $\Nf=1$ theory, scaled by the gradient flow scale $w_0$.}
\end{figure*}

\begin{figure}
    \includegraphics[width=\columnwidth]{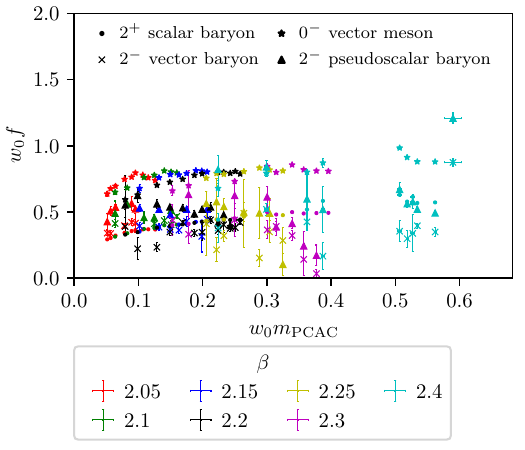}

    \caption{\label{fig:decayconst_Nf1}Spectrum of unrenormalised decay constants in the $\Nf=1$ theory, scaled by the gradient flow scale $w_0$.}
\end{figure}

We present the spectrum of masses and unrenormalised decay constants respectively
of the $\Nf=1$ theory,
including mesons, baryons, glueballs, the string tension, and the hybrid fermion,
in Figs.~\ref{fig:masses_Nf1} and~\ref{fig:decayconst_Nf1},
in terms of the gradient flow scale $w_0$.

\begin{figure}
    \includegraphics[width=\columnwidth]{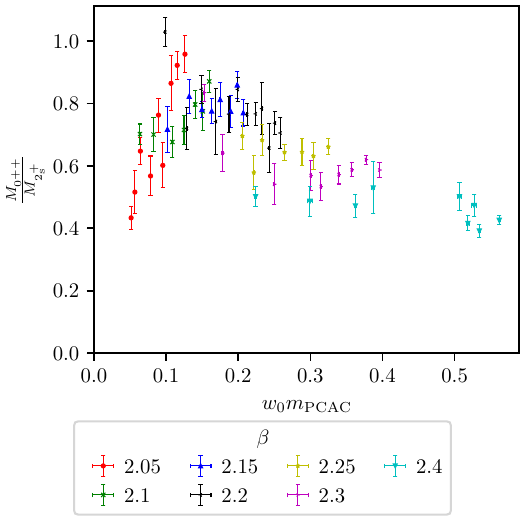}

    \caption{\label{fig:scalar_ratio}Ratio of the mass of the scalar glueball to the lightest fermionic bound state in the $\Nf=1$ theory.}
\end{figure}

\begin{figure}
    \includegraphics[width=\columnwidth]{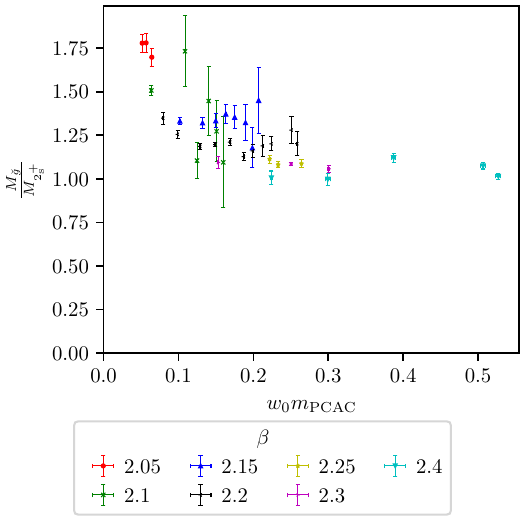}

    \caption{\label{fig:spin12_ratio}Ratio of the mass of the \spinhalf state to the lightest fermionic bound state in the $\Nf=1$ theory.}
\end{figure}

In particular,
we draw attention to two features of the spectrum.
The first,
highlighted in Fig.~\ref{fig:scalar_ratio},
is that,
as observed in our previous work
and in other theories in or near the conformal window,
the scalar glueball is lighter than the $2^+$ scalar baryon,
the would-be pion in a QCD-like theory.
Indeed,
this effect becomes more pronounced at higher values of $\beta$.
Secondly,
as can be seen in Fig.~\ref{fig:spin12_ratio},
in the $\Nf=2$ theory
the \spinhalf state is also lighter than the $2^+$ scalar baryon.

\subsection{Determination of the mass anomalous dimension}

\subsubsection{Finite-size hyperscaling of particle spectrum}

In a mass-deformed conformal theory in a finite volume, we expect the masses $M$ of states to hyperscale with the deforming mass (which we approximate by $\mpcac$) and the lattice extent $L$ as
\begin{equation}
	L a M = f \left(L\left(a\mpcac\right)^{\frac{1}{1+\gamma_*}} \right)\;.
	\label{eq:functiof}
\end{equation}

While we do not \emph{a priori} know the functional form of $f\left(L\left(a\mpcac\right)\right)$, we may look for a value of $\gamma_*$ that causes the data to collapse onto a single universal curve. As in Ref.~\cite{Athenodorou:2021wom}, we may make this determination more systematic by defining the curve collapse as the minimisation of the function
\begin{equation}
    P(\gamma_*) = \frac{1}{N_\mathrm{o}} \sum_{\ell \in \{L\}} \sum_{i\in S_{\ell}} \left(L_i aM_i - f_{\ell}\left(L_i^{1+\gamma_*} am_{\mathrm{PCAC}\,i} \right)\right)^2\;,
\end{equation}
where $\{L\}$ is the set of lattice extents, $S_{\ell}$ is the set of data having lattice extent $L_i \ne \ell$ but lying such that $L_i^{1+\gamma_*}$ is in the range spanned by data with lattice extent $\ell$, $N_{\mathrm{o}}$ is the number of points summed over, and $f_{\mathrm{\ell}}$ is an interpolating function across the data having lattice extent $\ell$, which we choose to be piecewise linear. We apply this procedure for the data for each value of $\beta$ in both $\Nf=1$ and $2$.

\begin{figure*}
    \includegraphics[width=\textwidth]{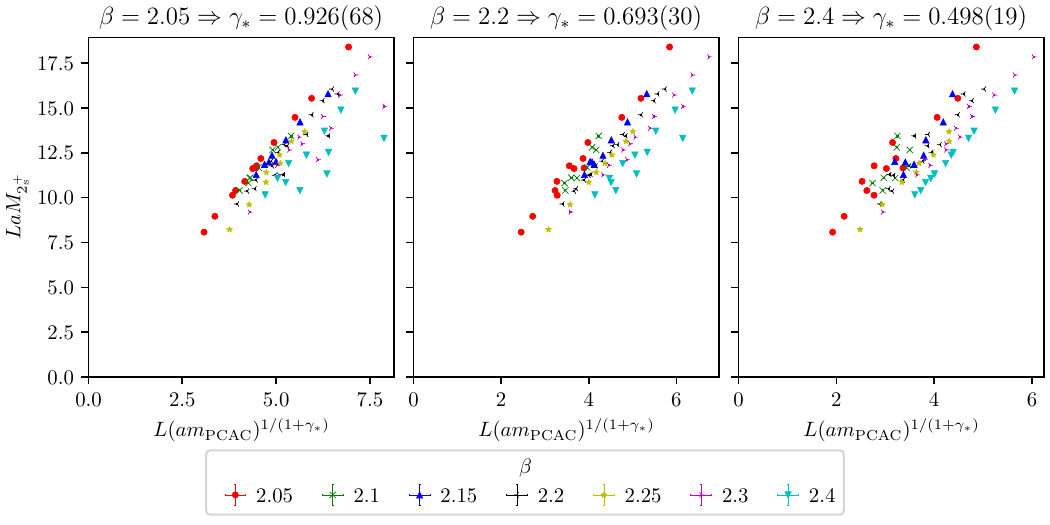}

    \caption{\label{fig:fshs_Nf1}Finite-volume hyperscaling fit results for $\Nf=1$. In each panel, the value of $\gamma_*$ used for the horizontal co-ordinate is that obtained from the FSHS fit of the data at a single value of $\beta$, as indicated above the plot. Data from the remaining values of $\beta$ are included in each case for comparison.}
\end{figure*}

\begin{table}
    \caption{\label{tab:gamma_Nf1}Results for the values for $\gamma_*$ obtained for $\Nf=1$ both by fitting the spectrum with a finite-size hyperscaling Ansatz, and by fitting the Dirac mode number. The mode number has not been fitted for the case $\beta=2.25$, due to the absence of larger-volume ensembles in that case.}

    \input{assets/tables/gamma_Nf1}
\end{table}

The results of this analysis are plotted for a selection of $\beta$ values considered in Fig.~\ref{fig:fshs_Nf1}
and the resulting values of $\gamma_*$ tabulated in Tab.~\ref{tab:gamma_Nf1}. As observed in our previous work~\cite{Athenodorou:2021wom}, we see the anomalous dimension in the $\Nf=1$ theory continue to decrease as $\beta$ increases. 

\subsubsection{$\gamma^*$ from the mode number of the Dirac operator}\label{sssec:modenu}
The other method we employ to calculate the mass anomalous dimension is studying the low-lying mode number of the Dirac operator. A similar computation to extract $\gamma^*$ was performed in previous publications~\cite{Athenodorou:2021wom,Athenodorou:2014eua}, to which we will refer to for additional details. However, the final analysis of the data has been improved and follows a slightly different strategy. For this reason, it is useful to review some general concepts, that will be useful in the following as they will guide the statistical analysis of the data.

The quantity of interest is the dependence of the spectral density $\rho$ of the Dirac operator on its eigenvalues $\omega$. In fact, in an infrared conformal gauge theory, at the IR fixed point, the leading behaviour of $\rho(\omega)$ is dictated by the renormalization group (RG) equations and proves to be power-like in $\omega$, where the anomalous dimension $\gamma^*$ appears as an exponent~\cite{Patella:2012da}. Equivalently, a more accessible quantity is the integral of the spectral density, called \textit{mode number}, i.e.~the number of eigenvalues of the Dirac operator lower than a certain mass threshold $\Omega^2$ (divided by the volume). As argued in Refs.~\cite{Patella:2012da,Cheng:2013eu}, on the lattice, the mode number in lattice units $a^{-4}\bar\nu$, can be expressed as a function of the threshold $a\Omega$ as
\begin{equation}
	a^{-4}\nubar(\Omega) \approx a^{-4} \nubar_0(m) + A[(a\Omega)^2-(am)^2]^{\frac{2}{1+\gamma_*}}\;.
\label{eq:nubar-scaling}
\end{equation}
where $a^{-4}\bar\nu_0$, $A$, $am$ and $\gamma^*$ are parameters to be fitted to the data. In Eq.~\eqref{eq:nubar-scaling}, the $\approx $ sign signals the fact that the equation is valid only where the dynamics is dominated by the IR fixed point, namely a region $\omega_\text{IR}<\omega<\omega_\text{UV}$ of low eigenvalues insensitive to the fermion mass. The parameters appearing in Eq.~\eqref{eq:nubar-scaling} are of physical relevance and contain information that can be incorporated into the analysis of the data. Specifically:
\begin{itemize}
    \item The overall additive constant $\bar \nu_0$ represents the integral of the spectral density in the region $[0,\omega_\text{IR}]$,
    where the fermion mass drives the theory away from the IR fixed points, spoiling scale invariance. In this region, no analytical models are available for $\rho$, and therefore its integral is represented by a constant $\bar{\nu}_0$. As the fermion mass is responsible for this behaviour, this region shrinks in the chiral limit, and, as argued in Ref.~\cite{Patella:2012da}, it can be parameterized as
    \begin{equation}\label{eq:nu0scaling}
        \bar\nu_0(m)\propto M^4
    \end{equation}
    where $M$ is some mass in the spectrum vanishing in the chiral limit.
    \item The mass parameter $m$ can be related to the bare PCAC mass as
    \begin{equation}\label{eq:numassscaling}
        m = Z_A \mpcac\,.
    \end{equation} 
\end{itemize}

The procedure for the extraction of the anomalous dimension consists of the following. First, the mode number $\bar\nu$ has to be computed through spectral projection and the Chebyshev expansion method as described in \cite{Bergner:2016hip}. We have confirmed consistency between both methods and the main part of the results has been obtained with the Chebyshev expansion. This provides the mode number for several values of the threshold $a\Omega$ on a subsample of well-decorrelated configurations. Then a fit of Eq.~\eqref{eq:nubar-scaling} to the data has to be performed to extract the value of the anomalous dimension $\gamma^*$. In doing so two main issues have to be addressed. 

The first issue is the choice of the window $[a\Omega_\text{min},a\Omega_\text{max}]$. This range is expected to reflect the region of validity of Eq.~\eqref{eq:nubar-scaling} and its location is not known \textit{a priori} and has to be determined empirically from the data. In principle, when choosing an inappropriate window, i.e.~a region where Eq.~\eqref{eq:nubar-scaling} is not valid, the low quality of the choice of $a\Omega_\text{min}$ and $a\Omega_\text{max}$ is expected to be signalled by a bad-$\chi^2$ fit. In practice, the usage of the simple $\chi^2$ as an indicator of the quality of the fit is not always a good discriminator. In fact, if the window is small enough and centred on high eigenvalues, the fit converges with a good $\chi^2$ to unphysical values of the parameters.\footnote{As pointed out in Ref.~\cite{Patella:2012da}, the spectral density has an asymptotic behaviour dominated by asymptotic freedom, given by $\rho\propto\omega^3$. If the fitting region is small enough, the fit of Eq.~\eqref{eq:nubar-scaling} to the data can give an acceptable $\chi^2$, leading to a vanishing value of $am$ and a $\gamma^*$ with no physical meaning.} For this reason, fits in a low-eigenvalues window are to be preferred, although a second following issue is expected to be triggered; the high degree of non-linearity of the function in Eq.~\eqref{eq:nubar-scaling} makes the fit highly unstable, the minimization algorithm often drives towards a region in the parameter space where $((a\Omega)^2 - (am)^2)<0$ and leads to a breakdown of the fitting procedure. In these cases, the fit parameters can be bounded or reparameterized to avoid this problem, although, we observed that in some cases the algorithm finds other local minima, and the parameters converge to extreme (unphysical) values. 

\begin{figure*}
    \centering
    \includegraphics[width=\textwidth]{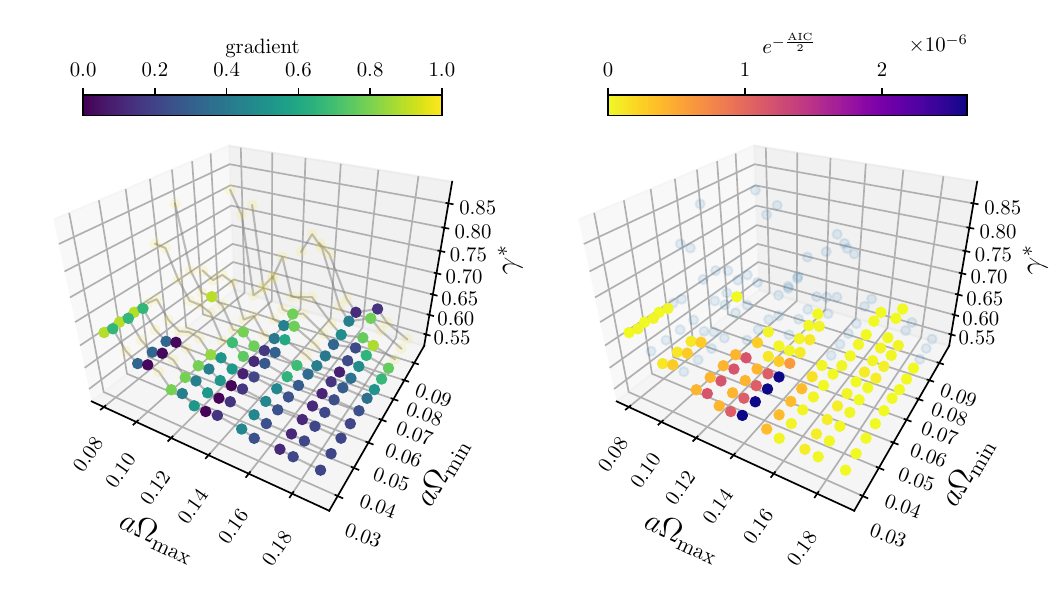}
    \caption{Values of the anomalous dimension $\gamma^*$ extracted from the fit for the ensemble \texttt{DB4M11}. Neighbouring points are joined by grey lines to guide the eye. \textit{Left panel}: the colour scale of the points is proportional to the gradient, colder colour corresponds to a flatter region. All the points whose gradient is lower than a certain cutoff value are depicted with a solid marker, the others with a transparent one. \textit{Right panel}: the colour scale of the point is proportional to the AIC weight given in Eq.~\eqref{eq:AIC}.  }
    \label{fig:nu_surface}
\end{figure*}

\begin{figure*}
    \centering
    \includegraphics[width=\textwidth]{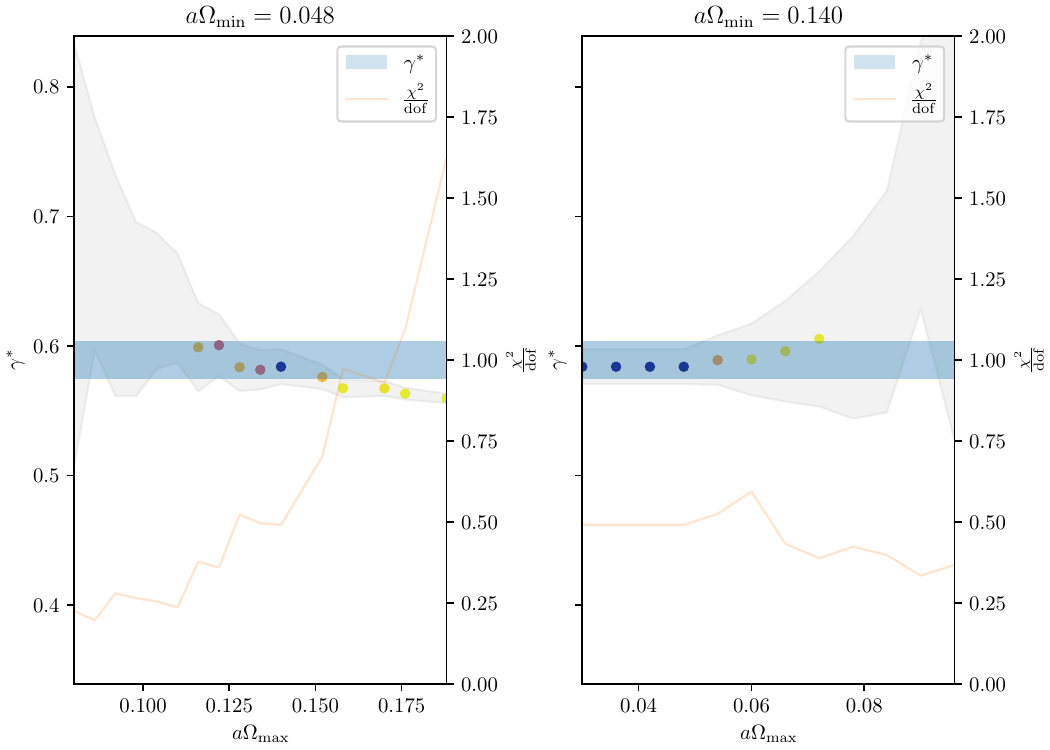}
    \caption{Values of the anomalous dimension $\gamma^*$ as a function of the higher (lower) end of the fitting window at a fixed value of lower (higher) end for the ensemble \texttt{DB4M11}. Error of single points are depicted as a continuous grey band, while point markers are coloured proportional to their model weight (calculated) with the same colour scale as in the right panel of Fig.~\ref{fig:nu_surface}. The weighted average is represented as a horizontal blue band whose size is proportional to its statistical error. The orange solid line corresponds to the normal (i.e.~not augmented) $\chi^2$, divided by the degrees of freedom.}
    \label{fig:nu_slice}
\end{figure*}

In previous publications~\cite{Athenodorou:2021wom,Athenodorou:2014eua}, these two issues were dealt with by repeating a series of bootstrapped fit to an extensive range of possible window $[a\Omega_\text{min},a\Omega_\text{max}]$ and thus seeking by eye for a region of stability. 
Coefficients were then extracted by averaging over the windows inside the stability region, where its fit is weighted with the relative portion of converged fit in a bootstrap sample, as an indicator of stability. In this work, we adopt a different approach, which reduces the number of arbitrary choices in the analysis and frames the procedure within a more solid statistical framework. To address the problem of the stability of the fit, we constrain fit parameters to physically reasonable ranges by means of Bayesian constraints with Gaussian priors~\cite{Lepage2002}. The procedure consists in the minimization of the \textit{augmented chi-square} $\chi_\text{aug}^2 = \chi^2 + \chi^2_\text{pr}$, defined as
\begin{equation}
    \chi^2(\mathbf{p}) = \sum_{i,j}[\nu_i-f_i(\mathbf{p})]C^{-1}_{ij}[\nu_j-f_j(\mathbf{p})]\,,
\end{equation}
where the index $i$ runs over all the data points within the fitting window, $\nu_i$ represents the mode number data computed on the lattice for a specific mass threshold $a\Omega_i$, $f_i(\mathbf{p})$ corresponds to the model function in Eq.~\eqref{eq:nubar-scaling} applied to a specific value of $a\Omega_i$ within the window for given values of the parameter vector $\mathbf{p}=[a^{-4}\bar\nu_0,A,am,\gamma^*]$, and $C$ is the full covariance matrix estimated from the Monte-Carlo data. In addition to that, we consider uncorrelated Gaussian priors for the parameters $\tilde p_n$, $\tilde \sigma_{p_n}$ that contribute to the global loss function as
\begin{equation}
    \chi^2_\text{pr}(\mathbf{p},\{
    \tilde p_n,\tilde\sigma_{p_n}
    \}) = \sum_{n=1}^4 \frac{(p_n-\tilde p_n)^2}{\tilde\sigma_{p_n}^2}\,.
\end{equation}
The selection of their central value is performed in a \textit{physics-informed} way. Following Eq.~\eqref{eq:numassscaling}, the prior of $am$ is centred to the value of $a\mpcac$ defined in Sec.~\ref{ssec:states_involving_fermions} times a coefficient which we set to 1.\footnote{This coefficient should correspond to the renormalization factor $Z_A$. We do not have a determination, but we assume it to be of order 1.} 
On the base of Eq.~\eqref{eq:nu0scaling}, the prior of $a^{-4}\bar\nu_0$ is centred to the lattice estimate of the (fourth power of the) scalar baryon mass times a coefficient of order 1. The priors for $A$ and $\gamma^*$ are set to $1(10)$ and $0.5(5.0)$, respectively.
It is to be stressed that priors are only meant to improve the convergence of the fit as they supposedly prevent the minimization algorithm from being stuck in some local minimum. The choice of their numerical values should not influence the final result. To ensure the validity of this assumption, as it is common in this Bayesian framework, the width of the prior should be made large enough to accommodate significant fluctuations from the central value, i.e.~choosing a so-called \textit{flat} prior. Following this criterion, the widths of the priors are set to account for a 1000\% relative error in the choice of the central values of $\gamma^*$ and $A$'s priors, 100\% for $am$ and $a^{-4}\nu_0$. In order to ensure the positivity of $a^{-4}\bar{\nu}_0$ and $am$, we use a log-normal distribution for their prior.
We repeat the constrained fit for a large number of possible windows by varying $a\Omega_\text{min}\in[0.03,0.1]$ and $a\Omega_\text{max}\in[0.08,0.18]$. 
As an example of the results of the fit, in Fig.~\ref{fig:nu_surface} we display the anomalous dimension $\gamma^*$ extracted from a fit, as a function of the parameters $a\Omega_\text{min}$ and $a\Omega_\text{max}$ for the ensemble \texttt{DB4M11} in a three-dimensional space with coordinates $(a\Omega_\text{max},a\Omega_\text{min},\gamma^*)$.
In order to seek a stable region in an automatized way, we calculate the gradient of the surface generated by our $\gamma^*$ value and we filter those points in which the norm of the gradient is below a certain threshold, in order to automatize the identification of the stable region. In the upper panel of Fig.~\ref{fig:nu_surface} we assign a colder colour to those points whose gradient is smaller, i.e.~corresponding to a ``flatter'' and therefore stabler region. The final result has to be given by averaging the coefficients over the different fitting windows in the stable region. This last scenario is analogous to the issue of \textit{data truncation} arising in other contexts such as hadron spectroscopy in lattice QCD. In the last few years, \textit{Bayesian model averaging} techniques to address this problem have started to raise a lot of interest~\cite{Jay:2020jkz, Frison:2023lwb, Borsanyi2015}, and have been used in several published results. Similar to our case, the goal is to address the problem of the selection of a subset of data for a fit and avoid being overly conservative in the estimation of the systematic error, by re-formulating the problem as a \textit{model selection} problem. 
Our proposal is to apply the methodology suggested in~\cite{Jay:2020jkz} by assigning to each window a so-called (modified) Akaike information criteria (AIC), defined as
\begin{equation}\label{eq:AIC}
    \text{AIC} = \chi_\text{aug}^2(\mathbf{p}^\star) + 2N_\text{cut}
\end{equation}
where $\chi_\text{aug}^2(\mathbf{p}^\star)$ is the augmented chi-square for the ``best-fit'' parameters $\mathbf{p}^\star$
and $N_\text{cut}$ is the number of available points minus the number of points in the fit.\footnote{In Ref.~\cite{Jay:2020jkz}, the weight factor differs for an extra factor $k$ in the exponent, $k$ being the number of parameters in the fitting function. Since in our case the number of fitted parameters is the same for every window, we drop it since it corresponds to an overall factor that disappears in the normalization.} A weight factor $w=e^{-\text{AIC}/2}$ is then assigned in the final average to each window in the stable region.
In order to illustrate how the different windows are weighted in the model average, in the right panel of Fig.~\ref{fig:nu_surface} we colour our points proportionally to their weight, calculated as in Eq.~\eqref{eq:AIC}. Using the same colour scale, in the left (right) hand side panel of Fig.~\ref{fig:nu_slice} we depict the value of $\gamma^*$ at fixed higher (lower) end of the window $a\Omega_\text{max}$ as a function of the lower (higher) one. As it is visible from the example, the model average tends to prefer bigger windows with good $\chi^2_\text{aug}$, as expected.
In Table~\ref{tab:gamma_Nf1} we report the values obtained for all the ensembles analyzed.

\subsubsection{$\gamma_*$ in the continuum limit}

\begin{figure}
    \includegraphics{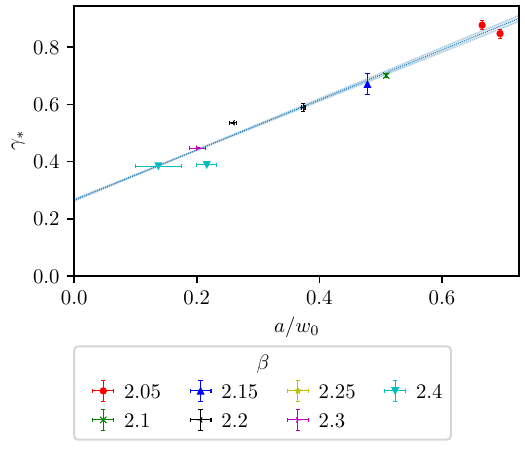}

    \caption{\label{fig:gamma-extrapolation-w0bare}
    The anomalous dimension $\gamma_*$ of the $\Nf=1$ theory
    obtained from the mode number
    plotted as a function of the per-ensemble value of the gradient flow scale $a / w_0$.
    Included for comparison is a fit form
    $\gamma_*(w_0) = \gamma_*^{\mathrm{cont.}}+c\frac{a}{w_0}$,
    which gives a continuum limit value $\gamma_*^{\mathrm{cont.}} = \GammaStarContinuumWZeroNfOne$.}
\end{figure}

We have previously observed,
and confirmed in the previous subsection,
that for the $\Nf=1$ case,
the value of $\gamma_*$ is not stable as the coupling $\beta$ is changed.
We have sufficiently many values of $\beta$
to attempt a continuum limit extrapolation.
Since the mode number method described above
allows computation of $\gamma_*$ for individual ensembles,
and there is a strong dependence of the gradient flow scale $w_0 / a$ on the fermion mass $am$,
one might expect to be able to study dependence of the per-ensemble $\gamma_*$
as a function of $a / w_0$,
and extrapolate to the $a / w_0 \rightarrow 0$ limit.
However,
as we show in Fig.~\ref{fig:gamma-extrapolation-w0bare},
$\gamma_*$ does not vary sufficiently with $am$ to align with
the trend observed as $\beta$ is changed.
Instead,
we must use a consistently-defined scale for each $\beta$ value.

\begin{figure}
    \includegraphics{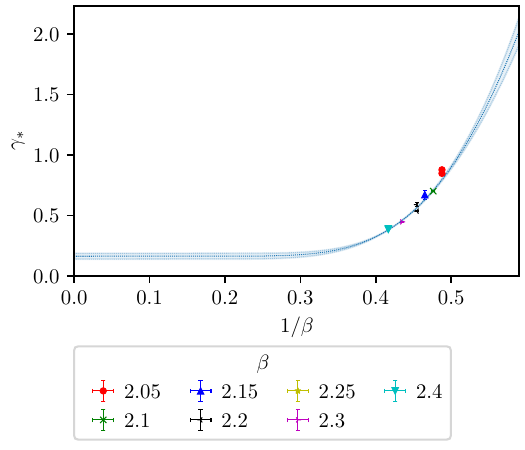}

    \caption{\label{fig:gamma-extrapolation-beta}
    The anomalous dimension $\gamma_*$ of the $\Nf=1$ theory
    obtained from the mode number
    plotted as a function of $\beta$.
    This is fitted using Eq.~\eqref{eq:gamma-contlim-fitform-beta},
    giving a continuum limit value $\gamma_*^{\mathrm{cont.}}=\GammaStarContinuumBetaLinearExponentNfOne$.}
\end{figure}

We adopt,
\emph{ad hoc},
two different prescriptions for this scale and the subsequent fitting form,
so that we can have some estimate of possible systematic effects.
Firstly,
we consider the behaviour of $\gamma_*$ obtained from the mode number
as a function of $1/\beta$ directly,
shown in Fig.~\ref{fig:gamma-extrapolation-beta}.
We observe that the data begin to flatten off as $\beta$ decreases.
We fit these data using the \emph{ad hoc} form
\begin{equation}
    \gamma_*(\beta) = d_0 + d_1 e^{ - d_2\beta} \;,
    \label{eq:gamma-contlim-fitform-beta}
\end{equation}
which is inspired by assuming a { leading power} correction in $a$,
with the latter parameterized with an exponential in $\beta$,
as suggested by the { scaling behaviour of the lattice spacing $a$
as a function of $\beta$ in the continuum limit (see, e.g.,~\cite{Boyd:1996bx,Lucini:2005vg} for similar approaches in the literature)}. 
This gives a continuum limit value of
$\gamma_* = \GammaStarContinuumBetaLinearExponentNfOne$,
with
$\chi^2/\mathrm{dof} = \GammaStarContinuumBetaLinearExponentNfOneExtrapolationChisquare$.
The high value of the $\chi^2/\mathrm{dof}$, which is mostly due to the lowest $\beta$ point, suggests that we are still too far from the continuum limit for Eq.~\eqref{eq:gamma-contlim-fitform-beta} to provide an accurate description of the data. { Indeed, taken at face value Fig.~\ref{fig:gamma-extrapolation-beta} would indicate that the asymptotic value of $\gamma_*$ would be reached for $\beta \ge 3.3$}. It is also possible that the errors on the extracted values of $\gamma_*$ are somewhat underestimated, or that an anaccounted systematic error on $\gamma_*$ affects our determination. Investigating the source of this discrepancy is beyond the scope of this work. However, we remark that adding a higher order term in the exponent reduces $\chi^2/\mathrm{dof}$ to around 7, which is still far from an acceptable value. 

\begin{figure*}
    \includegraphics{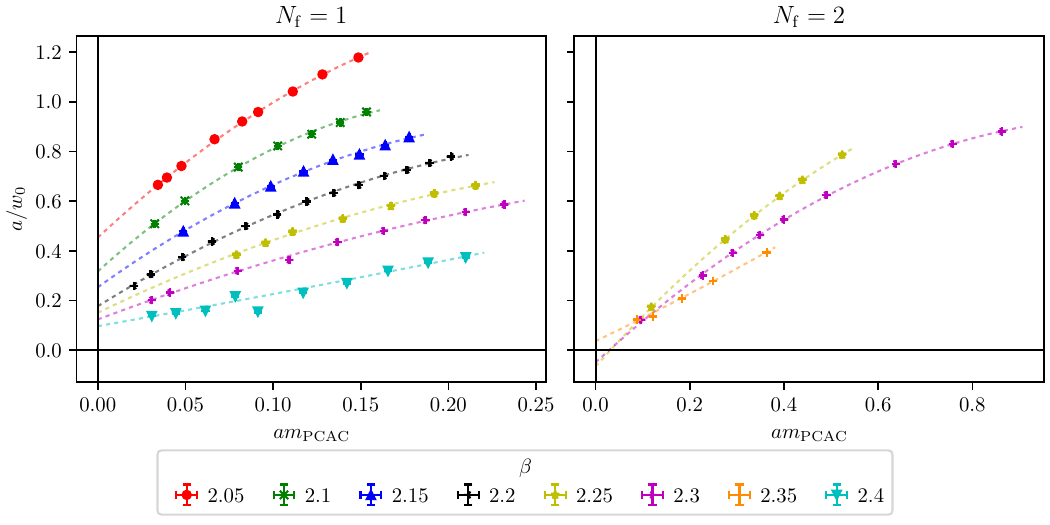}

    \caption{\label{fig:w0-extrapolation}
    The reciprocal gradient flow scale $a / w_0$ plotted as a function of $a\mpcac$
    for both $\Nf=1$ (left) and $2$ (right),
    including fits to Eq.~\eqref{eq:w0-fit-form} for each $\beta$ value.}
\end{figure*}

\begin{figure}
    \includegraphics{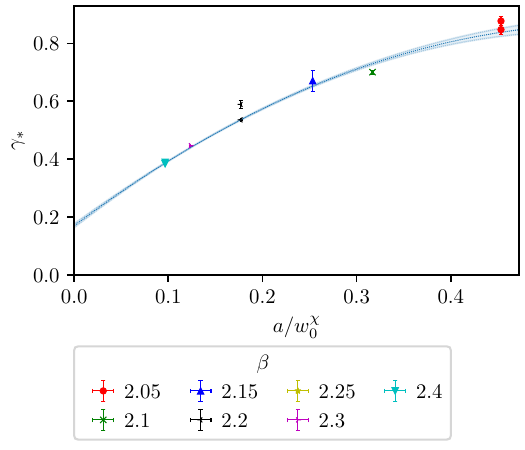}

    \caption{\label{fig:gamma-extrapolation-w0chiral}
    The anomalous dimension $\gamma_*$ of the $\Nf=1$ theory
    obtained from the mode number
    plotted as a function of the chiral limit reciprocal gradient flow scale $a/w_0^\chi$.
    This is fitted using Eq.~\eqref{eq:gamma-contlim-fitform-w0chiral},
    giving a continuum limit value $\gamma_*^{\mathrm{cont.}}=\GammaStarContinuumWZeroChiralNfOne$.}
\end{figure}

The second approach we consider is to take the chiral limit value for $a / w_0$.
This necessarily assumes that the theory is not in the conformal window,
where we would expect this value to be zero;
rather than infrared conformal, here we assume that the theory is near-conformal. We extrapolate $a / w_0$ as a function of $a\mpcac$ using a quadratic Ansatz,\footnote{{ This Ansatz originates from the chiral behaviour of quantities related to confinement, and includes the leading and sub-leading behaviours in $\mpcac$.}}
\begin{equation}
    \frac{a}{w_0} = \frac{a}{w_0^\chi} + B (a\mpcac) + C(a\mpcac)^2\;,
    \label{eq:w0-fit-form}
\end{equation}
where $a/w_0^\chi$ is the extrapolated chiral limit value of $a/w_0$.
This extrapolation is shown in Fig.~\ref{fig:w0-extrapolation}.
As $\beta$ increases,
$a/w_0^\chi$ approaches but does not reach zero, in agreement with our assumption of some residual global symmetry breaking. We then consider $\gamma_*$ as a function  of $a/w_0^\chi$, which we plot in Fig.~\ref{fig:gamma-extrapolation-w0chiral}.
We impose a quadratic Ansatz,
\begin{equation}
    \gamma_*(w_0^\chi) = \gamma_*^{\mathrm{cont.}} + c_0 \left(\frac{a}{w_0^\chi}\right) + c_1 \left(\frac{a}{w_0^\chi}\right)^2\;,
    \label{eq:gamma-contlim-fitform-w0chiral}
\end{equation}
which gives a resulting continuum limit of $\gamma_*^{\mathrm{cont.}} = \GammaStarContinuumWZeroChiralNfOne$.
As these data lie closer to the continuum limit,
the uncertainty on the extrapolated value is smaller
than the extrapolation in $1/\beta$ above.
(There may however be systematic uncertainties not accounted for
due to the choice of fit form.)
The result is consistent with the limit imposed by the fit to Eq.~\eqref{eq:gamma-contlim-fitform-beta}. Note however that in the case discussed here the value of $\gamma_*$ would be scheme dependent.

Taking the mean of these two estimates,
weighted by their statistical errors,
gives our final estimate for the continuum limit value
of the mass anomalous dimension of the $\Nf=1$ theory,
$\gamma_*=\GammaStarContinuumMeanNfOne$.


\subsection{Comparison with chiral perturbation theory}

In addition to,
and to prove a contrast with,
the hyperscaling case discussed above,
we also probe how well a
chiral perturbation theory Ansatz describes
the data in the infrared limit.
We fit the mass of the $2^+$ scalar baryon,
the pseudo-Nambu-Goldstone boson of the (potentially) broken chiral symmetry
{ (sourced by the $\gamma_5$ operator,
as for the pion in QCD)},
with the functional form:
\begin{align}
    \hat{M}_{2_{\mathrm{s}}^{+}} =&
    2B\hat{m}_{\mathrm{PCAC}}
    (1
    + L\hat{m}_{\mathrm{PCAC}}
    + D_1\hat{m}_{\mathrm{PCAC}}\log(D_2 \hat{m}_{\mathrm{PCAC}})) \nonumber\\
    &+ W_1 a\mpcac \nonumber\\
    &+ W_2 \frac{a^2}{w_0^2}\;,
    \label{eq:chipt}
\end{align}
where hatted quantities represent
the equivalent unhatted quantities
normalised by the gradient flow scale $w_0$.
We perform a simultaneous fit
across all values of $\beta$ for which data is available,
and then for each ensemble compute the fit value for $\hat{M}$
given the ensemble's observed values for
$\mpcac$ and $w_0$.

\begin{figure}
    \includegraphics[width=\columnwidth]{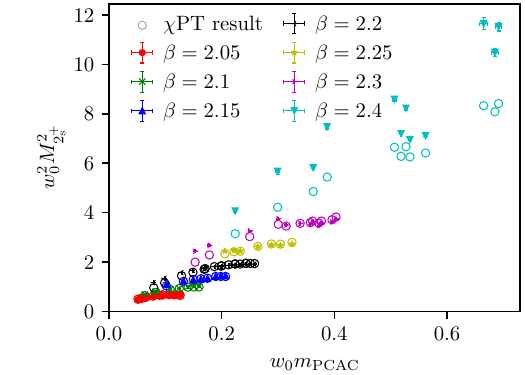}

    \caption{\label{fig:Xpt_Nf1}Comparison of the mass observed for the $2^+$ for the $\Nf=1$ theory {(sourced by the $\gamma_5$ operator; represented with error bars)} with a fit of these data to a chiral perturbation theory Ansatz (open circles).}
\end{figure}

The results for $\Nf=1$
are plotted in Fig.~\ref{fig:Xpt_Nf1}:
similarly to our previous work~\cite{Athenodorou:2021wom},
we see deviations from this form increasing
as the value of $a\mpcac$ decreases
or the value of $\beta$ increases.

\subsection{$R$ ratio}
\label{sec:R_ratio}
In this section, we examine our recently acquired data along with older results. Our aim is to explore universal characteristics that could delineate the broad parameter space of those strongly interacting gauge theories, in particular exposing their anomalous dimension as we approach the continuum. 

As shown in~\cite{Athenodorou:2016ndx},  in an appropriate regime, the ratio
\begin{eqnarray}
    R = \frac{M_{2^{++}}}{M_{0^{++}}} \, ,
\end{eqnarray}
where $M_{2^{++}}$ represents the mass of the lightest spin-2 composite state and $M_{0^{++}}$ denotes the mass of the lightest spin-0 state, has a universal behaviour that is determined by the value of the mass anomalous dimension $\gamma_*$, Therefore, the ratio $R$ serves as a useful tool for characterizing general properties within an infrared (IR) conformal gauge theory. Several noteworthy characteristics render the observable $R$ particularly significant. Crucially, $R$ exhibits scheme-independence, and remains unaffected by the internal global symmetries of the theory. Hence, the comparison of $R$ values computed in theories featuring vastly different internal symmetries and symmetry-breaking patterns provides an additional approach to the determination of $\gamma_*$ that proves especially valuable for gauge theories incorporating fermionic fields, where the intricacies of chiral symmetry and its breaking introduce non-trivial, model-dependent features.

Prior to presenting numerical results for $R$, let's briefly outline the anticipated behavior of a conformal theory under deformation by a fermion mass $m$. For a small, finite deforming mass region, spectral masses are expected to scale as $M \propto m^{1/\Delta}$ where $\Delta=1 + \gamma_{\star}$ represents the scaling dimension. Since a finite lattice volume imposes an absolute infrared cutoff, an appropriate scaling variable to consider on a finite box with typical dimension $L$ is $x = L m^{1/\Delta}$.  Consequently, the mass of a generic $i^{\rm th}$  state acquires a dependence on $x$, which can be expressed as $LM_i = f_i (x)$, with $f_i$ an \emph{a priori} unknown function. At leading order in $x$, for a state with mass $M_0$, this yields $LM_0 \propto x$, and through back substitution, $L M_i = f_i (LM_0) $.  Consequently, when considering mass ratios, the explicit dependence on the lattice extent cancels out, enabling comparison of data across different volumes:
\begin{eqnarray}
\frac{M_i}{M_j} = \frac{f_i(LM_0)}{f_j(LM_0)}\,.
\end{eqnarray}
We anticipate that the behavior of $R$ will exhibit four distinct regimes. For large values of $m$, $R$ tends to align closely with the value observed in pure \su{2} Yang-Mills theory, determined in lattice studies as $R = 1.44(4)$\cite{Lucini:2001ej, Lucini:2004my}. Conversely, at small $m$, the regime that is realised will hinge on the lattice size $L$. In sufficiently large volumes, the observed physics displays confining properties due to the mass deformation. Conversely, at very small volumes, the theory transitions to the so-called ``femto-universe" regime \cite{Athenodorou:2016ndx}, where previous studies suggest $R$ approaches unity~\cite{GonzalezArroyo:1987ycm, Daniel:1989kj, Daniel:1990iz}. However, for $L$ large enough and $m$ small enough, an intermediate regime of $R$ exists where conformal behavior can be observed. This intermediate region is of particular interest as in it $R$  can be extrapolated to obtain a chiral limit value that provides a universal signature of conformal behaviour. Note that whether the intermediate region is observed in a given simulation is a consequence of the choice of the lattice parameters, since a same value of $x$ can be obtained by tuning $m$ and $L$ to very different regimes.  

Given the universal behaviour of $R$, to gain insights on the anomalous dimension, we opt to contrast our findings with a toy model constructed using principles of gauge-gravity duality \cite{Athenodorou:2016ndx}. This calculable model, developed within the bottom-up approach to holography, incorporates only an operator with a scaling dimension $\Delta$ deforming the theory. Leveraging the obtained results for the anomalous dimension, we can compare the measured ratio with expectations derived from the toy model utilizing the corresponding value of $\Delta$. Further elucidation on this concept is available in Ref.~\cite{Athenodorou:2016ndx}.

\begin{figure}
    \includegraphics[height=0.9\textheight]{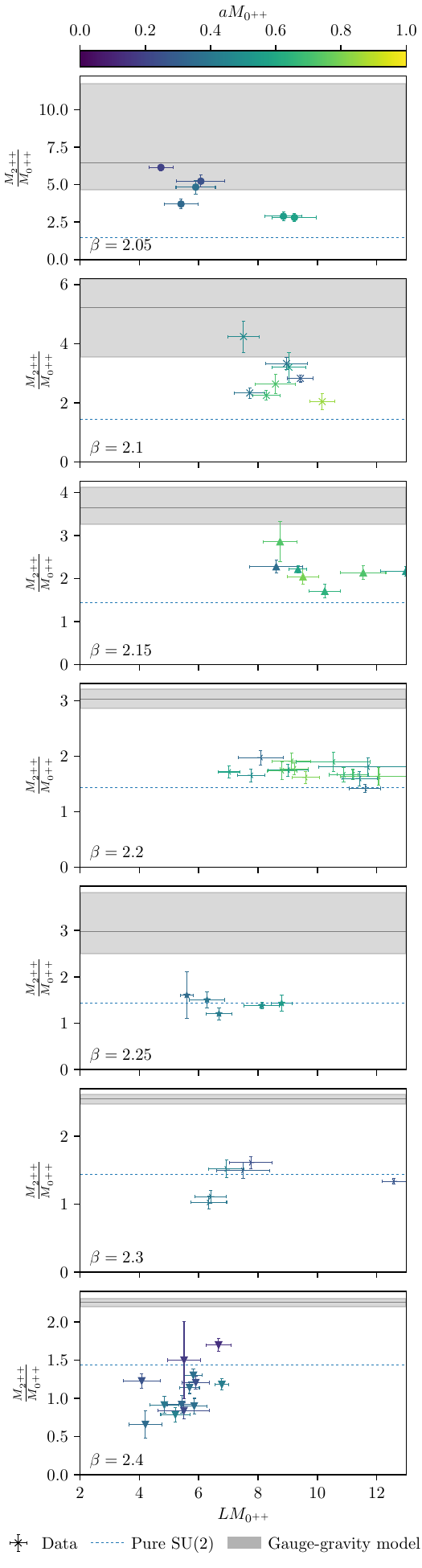}

    \caption{\label{fig:R_ratio_nf1}The value of the $R$ ratio observed for $\Nf=1$ for each value of $\beta$.}
\end{figure}

In Fig.~\ref{fig:R_ratio_nf1}, we display our findings regarding $R$ across seven distinct values of the coupling $\beta$ for \su{2} at $\Nf=1$. The data for the initial four plots have been previously showcased in Ref.~\cite{Athenodorou:2021wom}. Our results are juxtaposed with the $R$ ratio for the pure $\su{2}$ Yang-Mills theory (depicted by the blue dashed line) and with predictions derived from the gauge-gravity inspired toy model (shown as the upper band). It is worthy to note that the conformal prediction varies for each value of $\beta$ due to differing measured values of $\gamma_*$. Additionally, for $\beta=2.05$, the width of the band differs slightly from that presented in Ref.~\cite{Athenodorou:2016ndx} due to a different method of extracting $\gamma_*$ resulting in a more conservative estimation of the error: in Ref.~\cite{Athenodorou:2016ndx}, the results were derived using the Dirac mode number; in contrast, our current investigation employs the finite-size hyperscaling Ansatz.

For large values of $LM_{0^{++}}$, the data for $\beta=2.05$ up to $\beta=2.3$ tends toward the pure gauge prediction, in line with expectations. As $LM_{0^{++}}$ decreases, approaching the smallest values of $\beta$, the calculations become consistent with the conformal plateau, particularly for larger $R$. At $\beta=2.1$ and, to a lesser extent, $\beta=2.15$, there is some indication of behavior diverging from confinement, although the ratio does not rise sufficiently to match the value predicted by the gauge/gravity model. Notably, for the highest values of $\beta$, the results for $R$ do not significantly exceed the Yang-Mills value, even for large $LM_{0^{++}}$, possibly indicating an increasing difficulty in realising the regime that is relevant for near-conformality in this observable as $\beta$ increases. Interestingly, at the highest two values of $\beta$ and lower $LM_{0^{++}}$, there is clear evidence of $R$ decreasing towards $R \approx 1$, indicating a femto-universe behavior; In this region, the mass of the ground scalar glueball as well as the tensor glueball become approximately degenerate due to the lattice box geometry.

Overall we observe that for $\Nf=1$, the value of $R$ decreases from a scenario with a large anomalous dimension  ($R \gg 1$) as the value of $\beta$ increases and approaches the continuum limit. In fact, any signals of conformality become weaker and are eventually lost, which aligns with the idea that as the continuum is approached, the anomalous dimension decreases. This effect increses the difficulty in tuning the simulation to the intermediate region that is relevant for determining the anomalous dimension using the universality of $R$. Followingly, for large values of $LM_{0^{++}}$, the ratio $R$ converges to the pure Yang-Mills prediction of $R=1.44(4)$, enhancing the scenario of existing universality. Finally, for small values of $L M_{0^{++}}$ the $R$-ratio appears to be aligned with the femto-universe scenario.

Similarly, in Fig.~\ref{fig:R_ratio_nf2}, we present our results for the ratio $M_{2^{++}}/M_{0^{++}}$ at inverse couplings of $\beta=2.25$ and $\beta=2.35$. The findings for $\beta=2.25$ were previously discussed in detail in Ref.~\cite{Athenodorou:2016ndx}. Our current study extends these earlier results by including data for $\beta=2.35$, attempting for a more comprehensive analysis of the behavior of this mass ratio under varying coupling conditions.


Bearing the above in mind, let us attempt to understand our numerical findings. First, for large values of \(LM_{0^{++}}\), the data for \(\beta\) of 2.25 tends to align with the pure gauge prediction, as expected. As \(LM_{0^{++}}\) decreases, the calculations become consistent with the conformal plateau prediction; it appears that the latter is realised for values of \(R\) ranging for $LM_{0^{++}}=8 - 10$. This feature appears to weaken as we move closer to the continuum. Namely, for $\beta=2.35$ we cannot see an agreement with the conformal model anymore at intermediate values of $LM_{0^{++}}$ apart from one data-point for the very low values of $L M_{0^{++}}$ which appears to be an outlier. Instead, we observe that most of the data points are surrounding the pure gauge theory prediction. This suggests once more that the anomalous dimension weakens as we approach the continuum limit.

\section{Results for the $\Nf=2$ theory}
\label{sect:nf2}

In the following subsections we present the main results
obtained from studying the \su{2} theory with $\Nf=2$ flavours of adjoint fermion.
Compared to the previous work of Del Debbio et al.~\cite{DelDebbio:2008zf,DelDebbio:2010hx,Bursa:2011ru,Patella:2012da},
which studied a single value of the gauge coupling $\beta$,
we expand to study three values,
including replicating a subset of the parameters previously studied at $\beta=2.25$,
and moving to finer lattice spacing at $\beta=2.3$ and $2.35$,
as listed in Tab.~\ref{tab:ensembles-Nf2}.

\begin{table}
    \caption{\label{tab:ensembles-Nf2}
    Ensembles used for the $\Nf=2$ theory,
    showing the name of the ensemble,
    the coupling $\beta$,
    the bare fermion mass $am$,
    the number of sites in the temporal and spatial directions $N_t$ and $N_s$,
    {the total number of configurations considered $N_{\mathrm{conf.}}$,
    the number of trajectories between each saved configuration $\delta_{\mathrm{conf.}}$,
    and the length of the trajectory in molecular dynamics time units.}}
    \input{assets/tables/lattice_params_Nf2_part0}
\end{table}

As is the case for the $\Nf=1$ results above,
our main goal here is to better understand the dependence of observables on $\beta$.
In this section,
we only present the results and discuss their implications;
for further definitions and context,
we refer the reader to the previous section.

\subsection{Expectation value of Polyakov loops and center symmetry}

\input{assets/plots/polyakov_Nf2_caption}

Similarly to the $\Nf=1$ case,
we see no signs of a broken centre symmetry in the Polyakov loop distribution;
a selection are shown in Fig.~\ref{fig:polyakov-Nf2}.

\subsection{Topological aspects}

\input{assets/plots/q_topology_Nf2_caption}

In Fig.~\ref{fig:topcharge_Nf2} we show the history and histogram of the topological charge $Q$,
measured at Wilson flow time $\sqrt{8t_0}=\frac{L}{2a}$,
for the lightest-mass ensemble at each value of $\beta$.
For $\beta=2.25$,
we observe relatively rapid movement of $Q$,
albeit with some noticeable longer-autocorrelation modes;
the histogram is well fitted by a gaussian centered at $Q=0$,
as expected for an ergodic ensemble.
For $\beta=2.3$,
$Q$ moves much less ergodically,
with very long periods in which it is stationary;
however,
its histogram is still consistent with a narrow Gaussian centered at $Q=0$.
In the $\beta=2.35$ case,
the motion of $Q$ is much slower,
and its histogram is a long way from zero;
results from the lighter $\beta=2.35$ ensembles may be affected by
artefacts emerging from this lack of ergodicity in $Q$.

{ The topological susceptibility,
shown in units of the string tension $\sqrt{\sigma}$
as a function of the string tension in lattice units
in Fig.~\ref{fig:suscept},
shows close agreement with the equivalent results for the quenched theory;
following the argument in Sec.~\ref{sec:topology},
this is indicative of
a theory lying in the conformal window with small anomalous dimension.
For $\beta=2.3$, 2.35,
some deviation is seen at the lightest values of the fermion mass
(i.e. smallest value of the string tension);
as for the $\Nf=1$ case,
we attribute this to the effects of critical slowing down,
and include these data in the plot only to illustrate the difficulty.}

\subsection{Mass spectrum}

\begin{figure*}
    \includegraphics[width=\textwidth]{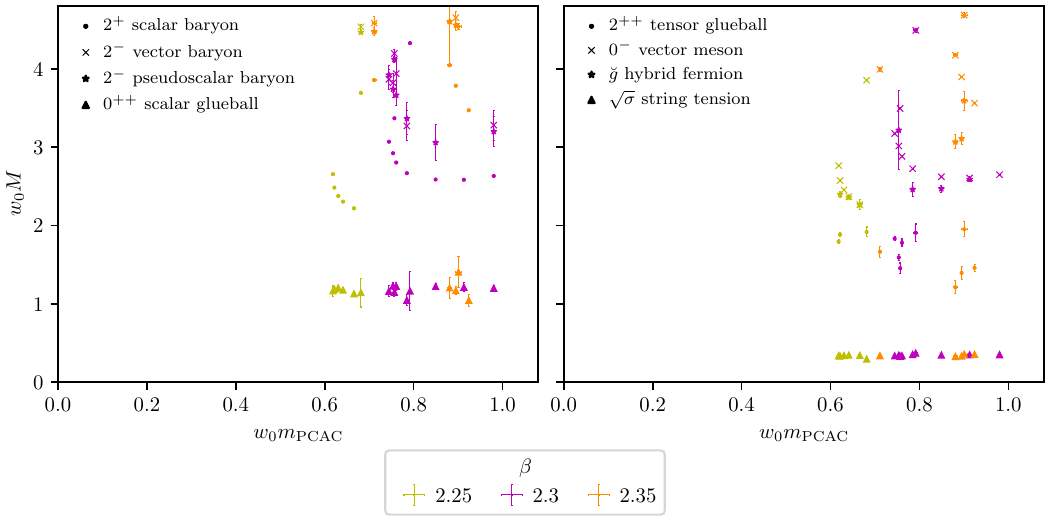}

    \caption{\label{fig:masses_Nf2}Spectrum of particle masses in the $\Nf=2$ theory, scaled by the gradient flow scale $w_0$.}
\end{figure*}

\begin{figure}
    \includegraphics[width=\columnwidth]{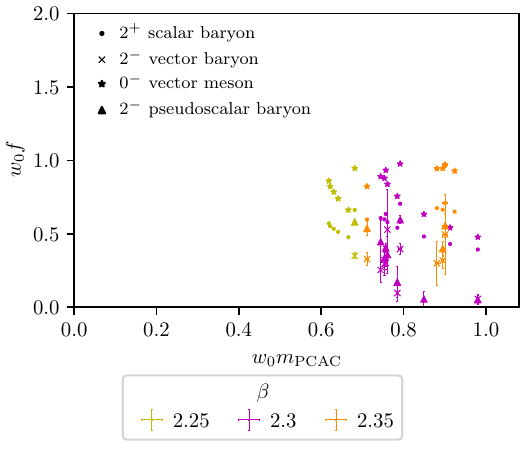}

    \caption{\label{fig:decayconst_Nf2}Spectrum of unrenormalised decay constants in the $\Nf=2$ theory, scaled by the gradient flow scale $w_0$.}
\end{figure}

In Fig.~\ref{fig:masses_Nf2},
we show the mass spectrum of the $\Nf=2$ theory as a function of the PCAC mass $\mpcac$,
both in terms of the gradient flow $w_0$.
As observed in the $\Nf=1$ theory in Fig.~\ref{fig:masses_Nf1},
we see also here that the string tension $\sqrt{\sigma}$ is a constant multiple of $w_0$.
We observe a trend towards increased mass of most states in units of $w_0$ at lighter values of the fermion mass,
visible most clearly in the two smallest values of $\beta$;
however,
the scalar glueball remains strikingly constant,
and consistent across all three $\beta$ values.
The corresponding decay constants are shown in units of $w_0$ in Fig.~\ref{fig:decayconst_Nf2};
unlike the $\Nf=1$ case,
we see neither flat behaviour at large mass,
nor a fall-off at small mass---instead,
we observe a gradual rise as the chiral limit is approached.

\begin{figure}
    \includegraphics[width=\columnwidth]{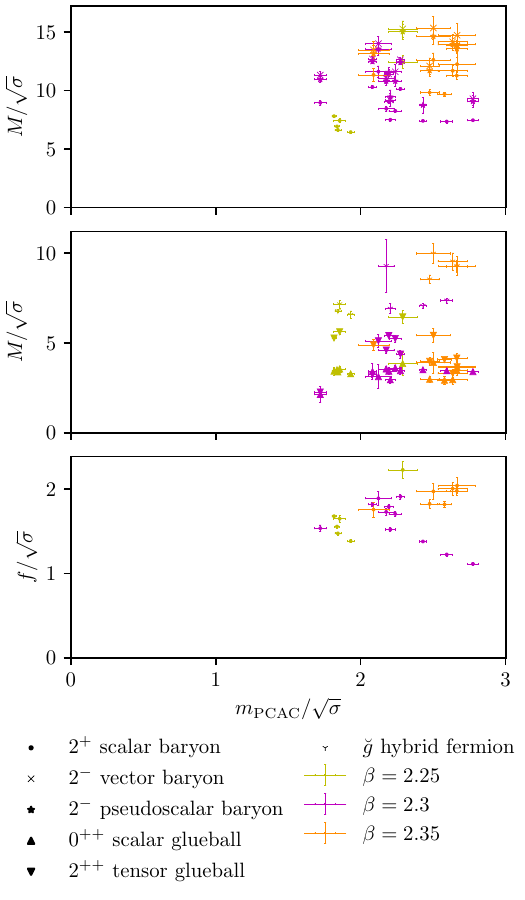}

    \caption{\label{fig:spectrum_sqrtsigma_Nf2}Spectrum of particle masses and (unrenormalised) decay constants in the $\Nf=2$ theory, scaled by the string tension $\sqrt{\sigma}$.}    
\end{figure}

We show in Fig.~\ref{fig:spectrum_sqrtsigma_Nf2}
the same quantities,
but expressed in terms of the string tension $\sqrt{\sigma}$
rather than the gradient flow $w_0$;
this allows showing additional ensembles for $\beta=2.35$
where $w_0$ could not be computed due to the finite lattice extent.
The same trends are visible as in the plots scaled by $w_0$,
but with larger statistical uncertainty due to the larger error on $\sqrt{\sigma}$.

\begin{figure}
    \includegraphics[width=\columnwidth]{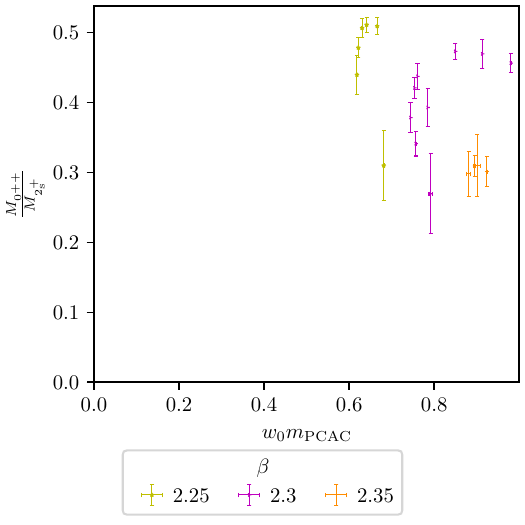}

    \caption{\label{fig:scalar_ratio_Nf2}Ratio of the mass of the scalar glueball to the lightest fermionic bound state in the $\Nf=2$ theory.}
\end{figure}

\begin{figure}
    \includegraphics[width=\columnwidth]{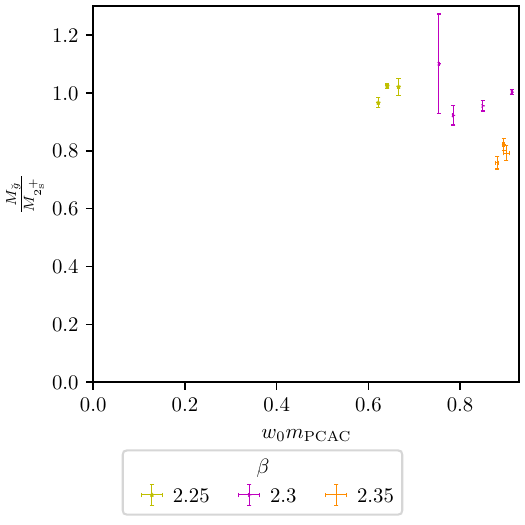}

    \caption{\label{fig:spin12_ratio_Nf2}Ratio of the mass of the \spinhalf state to that of the lightest fermionic bound state in the $\Nf=2$ theory.}
\end{figure}

Similarly to the above,
we highlight the ratio of the masses of two specific states to that of the $2^+$ scalar baryon
(the would-be PNGB):
the $0^{++}$ scalar glueball in Fig.~\ref{fig:scalar_ratio_Nf2},
and the \spinhalf state in Fig.~\ref{fig:spin12_ratio_Nf2}.
The former is significantly lighter than the $2^+$ scalar baryon
at all values of the coupling and the fermion mass;
the latter is comparable to the $2^+$ scalar baryon at small $\beta$,
becoming lighter than it at larger $\beta$.

\subsection{Mass anomalous dimension}

\begin{figure*}
    \includegraphics[width=\textwidth]{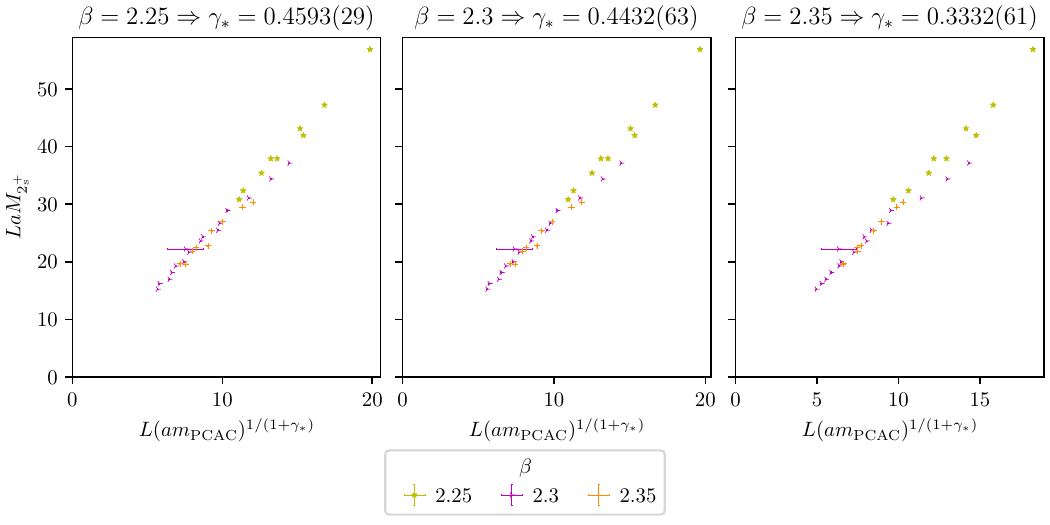}

    \caption{\label{fig:fshs_Nf2}Finite-volume hyperscaling fit results for $\Nf=2, \beta=2.35$.}
\end{figure*}

\begin{figure*}
    \centering
    \includegraphics[width=\textwidth]{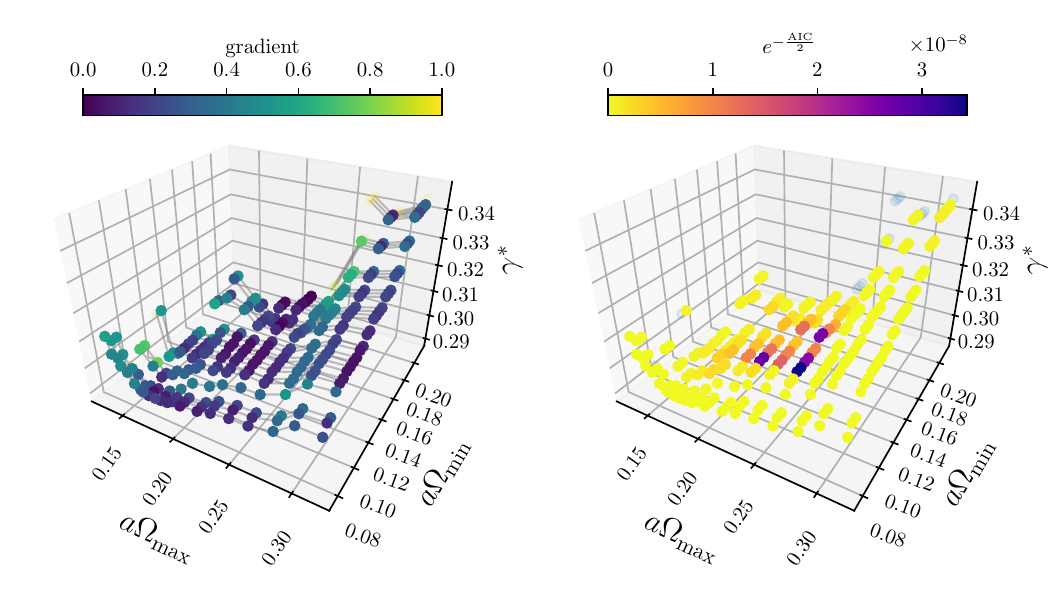}
    \caption{Values of the anomalous dimension $\gamma^*$ extracted from the fit for the ensemble \texttt{Nf2DB1M13}. Neighbouring points are joined by grey lines to guide the eye. \textit{Left panel}: the colour scale of the points is proportional to the gradient, colder colour corresponds to a flatter region. All the points whose gradient is lower than a certain cutoff value are depicted with a solid marker, the others with a transparent one. \textit{Right panel}: the colour scale of the point is proportional to the AIC weight given in Eq.~\eqref{eq:AIC}.  }
    \label{fig:nu_surface_Nf2}
\end{figure*}

\begin{figure*}
    \centering
    \includegraphics[width=\textwidth]{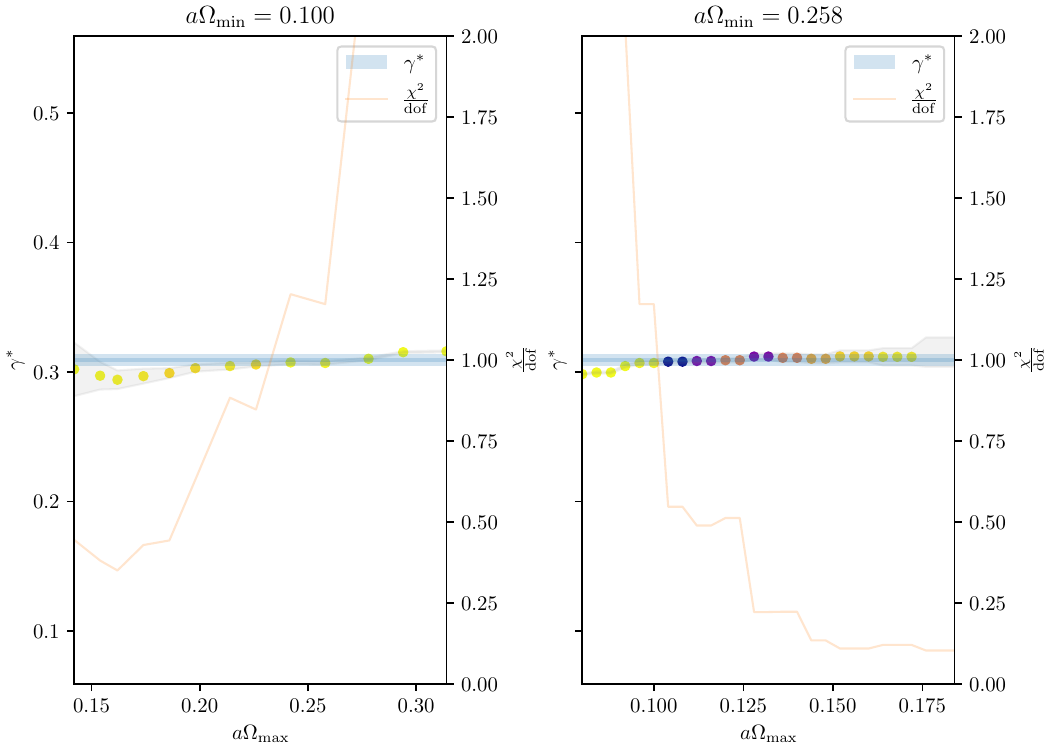}
    \caption{Values of the anomalous dimension $\gamma^*$ as a function of the higher (lower) end of the fitting window at a fixed value of lower (higher) end for the ensemble \texttt{Nf2DB1M13}. Error of single points are depicted as a continuous grey band, while point markers are coloured proportional to their model weight (calculated) with the same colour scale as in the right panel of Fig.~\ref{fig:nu_surface_Nf2}. The weighted average is represented as a horizontal blue band whose size is proportional to its statistical error. The orange solid line corresponds to the normal (i.e.~not augmented) $\chi^2$, divided by the degrees of freedom.}
    \label{fig:nu_slice_Nf2}
\end{figure*}

\begin{table}
    \caption{\label{tab:gamma_Nf2}Results for the values for $\gamma_*$ obtained for $\Nf=2$ both by fitting the spectrum with a finite-size hyperscaling Ansatz, and by fitting the Dirac mode number.}

    \input{assets/tables/gamma_Nf2}
\end{table}

We estimate the mass anomalous dimension via two routes:
a finite-size hyperscaling (FSHS) analysis for each value of $\beta$,
and a Bayesian fit of the Dirac mode number distribution for specific ensembles,
for the lightest ensemble for each value of $\beta$.
The results of the FSHS analysis for the $\Nf=2$ theory are shown in Fig.~\ref{fig:fshs_Nf2},
and an example mode number fit is illustrated in Figs.~\ref{fig:nu_surface_Nf2} and~\ref{fig:nu_slice_Nf2}.
The numbers are summarised in Tab.~\ref{tab:gamma_Nf2}.

Similarly to observations in the $\Nf=1$ theory,
the numbers show a similar trend,
but the values of $\gamma_*$ are
systematically lower for the mode number analysis
than for the FSHS analysis.
They do,
however,
become closer together at higher $\beta$.
Unlike the $\Nf=1$ theory,
there appears to be a rapid convergence of the mode number results.

\begin{figure}
    \includegraphics{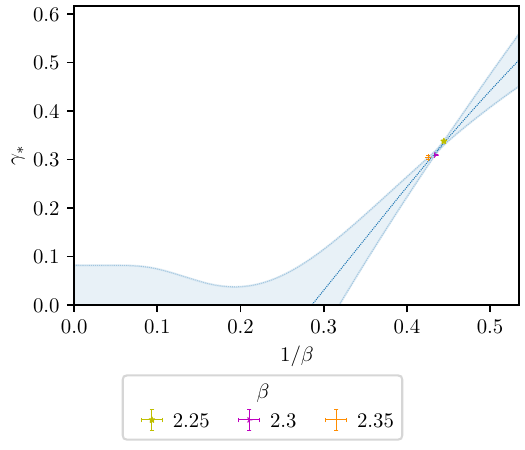}

    \caption{\label{fig:gamma-extrapolation-beta-Nf2}
    The anomalous dimension $\gamma_*$ of the $\Nf=2$ theory
    obtained from the mode number
    plotted as a function of $\beta$.
    For illustration,
    a fit using Eq.~\eqref{eq:gamma-contlim-fitform-beta} is shown;
    this fails to encompass the curvature of the data
    and the stability of the finest two lattice spacings.}
\end{figure}

\begin{figure}
    \includegraphics{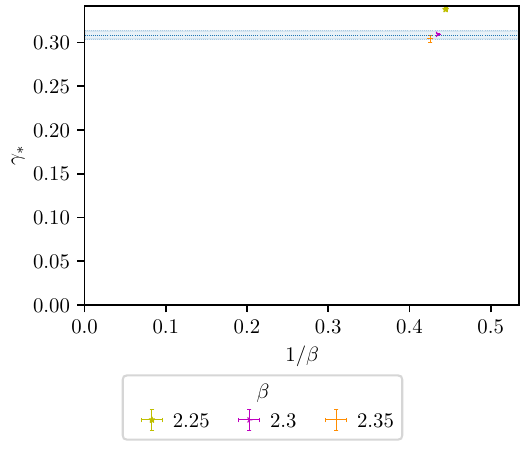}

    \caption{\label{fig:gamma-extrapolation-const-Nf2}
    The anomalous dimension $\gamma_*$ of the $\Nf=2$ theory
    obtained from the mode number
    plotted as a function of $\beta$.
    The values for the finest two lattice spacings are fitted with a constant,
    giving the value $\gamma_*=\GammaStarContinuumFlatNfTwo$.}
\end{figure}

The chiral limit value of $a/w_0$ for the $\Nf=2$ theory is very close to zero,
as can be seen in the right panel of Fig.~\ref{fig:w0-extrapolation},
so we cannot extrapolate as a function of this.
The fit form we chose \emph{ad hoc}
for $\gamma_*$ as a function of $\beta$
in the $\Nf=1$ theory does not correctly capture the curvature of the data in the $\Nf=2$ case,
as we illustrate in Fig.~\ref{fig:gamma-extrapolation-beta-Nf2}.
However,
as the data for the largest two $\beta$ values are
consistent within uncertainties,
we choose to take the weighted mean of these two values,
and estimate the continuum limit value as $\gamma_*=\GammaStarContinuumFlatNfTwo$,
illustrated in Fig.~\ref{fig:gamma-extrapolation-const-Nf2}.
It is clear from the figure that the extrapolation region is long compared to the spread of the data:
while the two points considered are consistent within uncertainties,
it is possible that this is only a metastability
or a slower-moving value;
this gives a non-negligible systematic uncertainty on the continuum limit value for $\gamma_*$ that we do not estimate here.

{ Our results have to be compared with other studies of the theory as summarized for example in \cite{Bergner:2016hip}. In these studies different methods have been applied leading to different systematic effects, but the range of precise estimates is roughly between $0.20$ and $0.38$. Hence there might be still a more significant scaling of the value than suggested by our limited data.}

\subsection{Comparison with chiral perturbation theory}

\begin{figure}
    \includegraphics[width=\columnwidth]{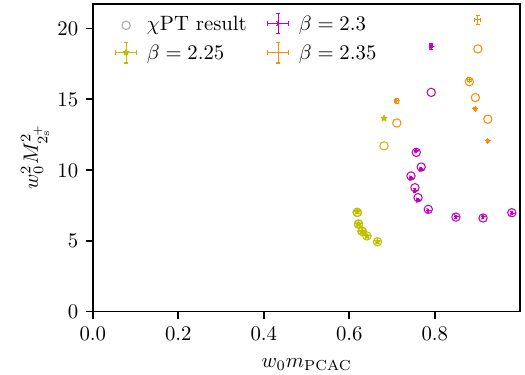}

    \caption{\label{fig:Xpt_Nf2}Comparison of the mass observed for the $2^+$ for the $\Nf=2$ theory {(sourced by the $\gamma_5$ operator; represented with error bars)} with a fit of these data to a chiral perturbation theory Ansatz (open circles).}
\end{figure}

{
As above for the $\Nf=1$ theory,
we check the compatibility of the $\Nf=2$ theory with chiral perturbation
by fitting the mass of the $2^+$ scalar baryon
(sourced by the $\gamma_5$ operator,
as for the pion in QCD)
as a function of the PCAC mass,
using Eq.~\eqref{eq:chipt}.}
As shown in Fig.~\ref{fig:Xpt_Nf2},
similarly to the former case,
the data for $\Nf=2$ are well described by chiral perturbation theory
at small $\beta$ and heavy mass,
but diverge from it for lighter masses and finer lattice spacings.

\subsection{$R$ ratio}

\begin{figure}
    \includegraphics[width=\columnwidth]{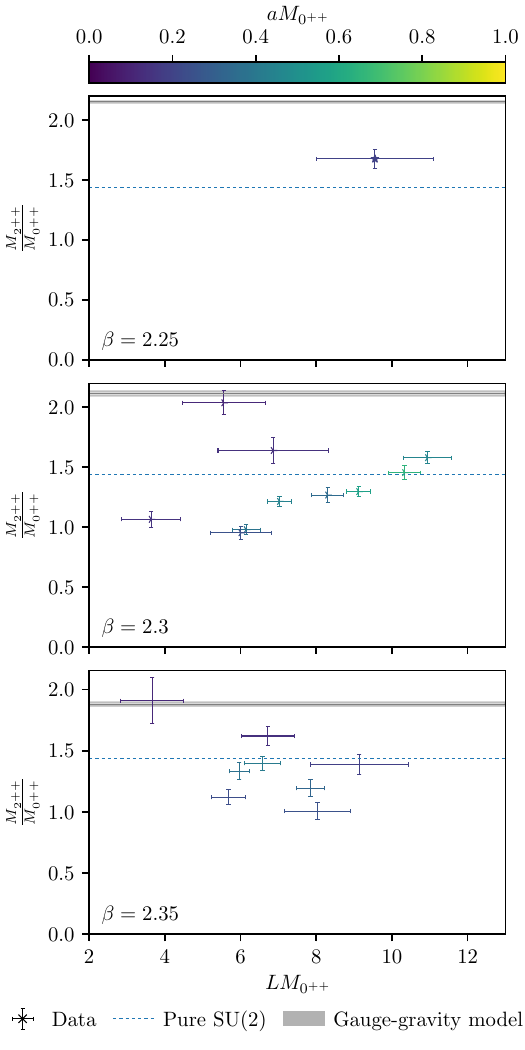}

    \caption{\label{fig:R_ratio_nf2}The value of the $R$ ratio observed for $\Nf=2$, $\beta=2.25,2.3,2.35$.}
\end{figure}

Figure~\ref{fig:R_ratio_nf2} shows
the ratio of the mass of the tensor to that of the scalar glueball.
Most heavier masses are consistent with the pure gauge result,
or sit below it,
potentially indicating some degree of finite-volume effects;
the lightest mass results
(darker colours)
however,
become consistent with the predictions of gauge-gravity duality~\cite{Athenodorou:2016ndx}
for the observed value of the mass anomalous dimension.

\section{Conclusions}
\label{sect:con}
Armed with the purpose of advancing the characterisation of the infrared behaviour of $\su{2}$ gauge theories with $\Nf = 1$ and $\Nf = 2$ Dirac fermion flavours transforming in the adjoint representation, we have performed an extended study of the former model in bare parameter space, with the intent of extrapolating the results to the continuum limit. For the considered values of the parameters, we have provided a detailed study of the mass spectrum, which has enabled us to extract the fermion condensate anomalous dimension using the expected hyperscaling behaviour for (near-)conformal systems. In addition, the anomalous dimension has been investigated making use of a mode number analysis of the Dirac spectrum and of the universal behaviour of the ratio of the mass of the lightest spin-two particle over the mass of the scalar. With different efficacy, all methods give consistent results and point towards heavy corrections to the anomalous dimension as the continuum limit is approached. Using {\em ad hoc} Ans\"atze, we have extrapolated the anomalous dimension to zero lattice spacing. Neglecting the systematics, including the one coming from the extrapolation, which can be somewhat large, our final result for the $\Nf = 1$ theory is $\gamma_* = \GammaStarContinuumMeanNfOne$. In contrast with earlier findings at larger lattice spacings, this value is significantly smaller than one.  Barring very large lattice artefacts, a scenario that would need to be investigated further, the strong dependence of the anomalous dimension on the lattice spacing seems to exclude a clear signature of infrared conformality. { However, we note that, as we increase $\beta$, the range of masses that we are able to access shifts towards higher values, affecting the robustness of the determination of the anomalous dimension in the chiral limit. In particular, a determination that uses higher masses is expected to result in a lower value of $\gamma_*$. The effect of the resulting systematics on our continuum extrapolation would need to be studied more in detail.} We have also performed a chiral perturbation theory analysis, which shows that a regime with detectable chiral symmetry breaking is also excluded. This however does not exclude the possibility that the infrared dynamics of the model will result in a small breaking of chiral symmetry. Putting together both the near-conformal and the chiral perturbation theory analysis, the most likely conclusion is that the theory is in the chiral symmetry broken phase, but rather close to the sill of conformality.

For the $\Nf = 2$ theory,
we have obtained results at three values of $\beta$,
one reproducing previous work and two larger values.
Combining the corresponding data with an analysis that parallels the $\Nf = 1$ case,
within the limitations of the information that can be extracted,
we note a milder dependence of $\gamma_*$ on the lattice spacing,
giving the appearance of a plateau at $\gamma_*=\GammaStarContinuumMeanNfTwo$,
and again albeit milder an intrinsic difficulty of chiral perturbation theory to describe the results.
A likely conclusion is that the $\Nf = 2$ model lies inside the conformal window, and has a small anomalous dimension. { It is also possible that in this case the (milder) dependency on the lattice spacing is at least in part due to the systematics related to the increasing difficulty to reach lower masses as $\beta$ is increased.}

After over 15 years of chasing the infrared behaviour of those models, this very important quest appears to be far from over. However, our work has pushed the boundaries of what can be done with state of the art resources using a few millions hours of A100 GPUs with the formulation of the action adopted so far. Future investigations call for the use of actions that have smaller lattice spacing corrections, can hence produce more physical results at a relatively coarse lattice spacing, and are more amenable to be pushed towards the chiral limit. To carry on this line of investigation, the possibilities that we are currently exploring include M\"obius domain wall fermions (see, e.g.,~\cite{Brower:2012vk}) and Pauli-Villars bosons~\cite{Hasenfratz:2021zsl}.  

\begin{acknowledgments}
{ We would like to thank M. Piai for discussions, in particular on high-mass systematics affecting the extrapolation of the anomalous dimension to the continuum limit.} 

A.A. was supported by the Horizon 2020 European research infrastructures
programme ``NI4OS-Europe” with grant agreement no. 85764, by ``SimEA" project funded by the European Union’s Horizon 2020 research and innovation programme under grant agreement No 810660, as well as by the ``EuroCC" project funded by the ``Deputy Ministry of Research, Innovation and Digital Policy and the Cyprus Research and Innovation Foundation" as well as by the European High-Performance Computing Joint Undertaking (JU) under grant agreement No.~101101903. 

P.B. acknowledges support by the project H2020-MSCAITN-2018- 813942 (EuroPLEx) and the EU Horizon 2020 research and innovation programme and by the Grant DGA-FSE grant 2020-E21-17R Aragon Government and the European Union - NextGenerationEU Recovery and Resilience Program on ``Astrofísica y Física de Altas Energías" CEFCA-CAPA-ITAINNOVA. 

The work of E.B. and J.L. has been supported by the UKRI Science and Technology Facilities Council (STFC) Research Software Engineering Fellowship EP/V052489/1. The work of E.B., J.L., and B.L. has been supported in part by the EPSRC ExCALIBUR programme ExaTEPP (project EP/X017168/1). E.B. and B.L. have been partly supported by the STFC  Consolidated Grant No. ST/T000813/1. B.L. received funding from the European Research Council (ERC) under the European Union’s Horizon 2020 research and innovation program under Grant Agreement No.~813942. 

G.B.\ is funded by the Deutsche Forschungsgemeinschaft (DFG) under Grant No.~432299911 and 431842497.

This work used the DiRAC Extreme Scaling service Tursa at the University of Edinburgh, managed by the Edinburgh Parallel Computing Centre on behalf of the STFC DiRAC HPC Facility (www.dirac.ac.uk). The DiRAC service at Edinburgh was funded by BEIS, UKRI and STFC capital funding and STFC operations grants. DiRAC is part of the UKRI Digital Research Infrastructure. We acknowledge the support of the Supercomputing Wales project, which is part-funded by the European Regional Development Fund (ERDF) via Welsh Government.

\end{acknowledgments}

\paragraph*{Open access statement}
For the purpose of open access, the authors have applied a Creative Commons
Attribution (CC BY) licence to any Author Accepted Manuscript version arising.

\paragraph*{Research Data Access Statement}
The data generated for this manuscript can be downloaded from Ref.\cite{datapackage2},
in addition to data from previous work already available from Ref.~\cite{datapackage1}
and the analysis code from Ref.~\cite{analysiscode}.

\bibliography{newreferences}

\end{document}

%% file: assets/definitions/gammastar_continuumlimit.tex
\newcommand \GammaStarContinuumWZeroNfOne {0.2654(35)}

\newcommand \GammaStarContinuumBetaLinearExponentNfOne {0.163(24)}
\newcommand \GammaStarContinuumBetaLinearExponentNfOneExtrapolationChisquare {12.08}
\newcommand \GammaStarContinuumWZeroChiralNfOne {0.1702(66)}

\newcommand \GammaStarContinuumMeanNfOne {0.170(6)}

\newcommand \GammaStarContinuumMeanNfTwo {0.291(9)}
\newcommand \GammaStarContinuumFlatNfTwo {0.3084(47)}

%% file: assets/tables/lattice_params_Nf1_part0.tex
\begin{tabular}{c|ccc|ccc}
     & $\beta$ & $am$ & $N_t \times N_s^3$ & $N_{\rm conf.}$ & $\delta_{\rm conf.}$ & $\tau_{\rm traj.}$ \\
    \hline
    \hline
    DB1M1 & $2.05$ & $-1.475$ & $32 \times 16^3$ & $4000$ & $1$ & $1.0$ \\
    DB1M1* & $2.05$ & $-1.475$ & $24 \times 12^3$ & $4000$ & $1$ & $1.0$ \\
    DB1M2 & $2.05$ & $-1.49$ & $32 \times 16^3$ & $4000$ & $1$ & $1.0$ \\
    DB1M3 & $2.05$ & $-1.5$ & $24 \times 12^3$ & $4000$ & $1$ & $2.0$ \\
    DB1M4 & $2.05$ & $-1.51$ & $48 \times 24^3$ & $4000$ & $1$ & $4.0$ \\
    DB1M4* & $2.05$ & $-1.51$ & $32 \times 16^3$ & $4163$ & $1$ & $4.0$ \\
    DB1M4** & $2.05$ & $-1.51$ & $24 \times 12^3$ & $4013$ & $1$ & $1.0$ \\
    DB1M5 & $2.05$ & $-1.514$ & $32 \times 16^3$ & $4000$ & $1$ & --- \\
    DB1M6 & $2.05$ & $-1.519$ & $32 \times 16^3$ & $4000$ & $1$ & --- \\
    DB1M7 & $2.05$ & $-1.523$ & $48 \times 24^3$ & $4000$ & $1$ & $1.0$ \\
    DB1M7* & $2.05$ & $-1.523$ & $32 \times 16^3$ & $4000$ & $1$ & $1.0$ \\
    DB1M8 & $2.05$ & $-1.524$ & $48 \times 24^3$ & $4000$ & $1$ & $2.0$ \\
    DB1M8* & $2.05$ & $-1.524$ & $32 \times 16^3$ & $4006$ & $1$ & $4.0$ \\
    DB1M9 & $2.05$ & $-1.5246$ & $48 \times 24^3$ & $4000$ & $1$ & $4.0$ \\
    \hline
    DB2M1 & $2.1$ & $-1.42$ & $24 \times 12^3$ & $4000$ & $1$ & $2.0$ \\
    DB2M2 & $2.1$ & $-1.43$ & $24 \times 12^3$ & $4000$ & $1$ & $2.0$ \\
    DB2M3 & $2.1$ & $-1.44$ & $24 \times 12^3$ & $4000$ & $1$ & $2.0$ \\
    DB2M4 & $2.1$ & $-1.45$ & $32 \times 16^3$ & $4000$ & $1$ & $2.0$ \\
    DB2M5 & $2.1$ & $-1.46$ & $32 \times 16^3$ & $4000$ & $1$ & $2.0$ \\
    DB2M6 & $2.1$ & $-1.47$ & $48 \times 24^3$ & $4000$ & $1$ & $2.0$ \\
    DB2M6* & $2.1$ & $-1.47$ & $40 \times 20^3$ & $4000$ & $1$ & $2.0$ \\
    DB2M7 & $2.1$ & $-1.474$ & $64 \times 32^3$ & $4000$ & $1$ & $4.0$ \\
    \hline
    DB3M1 & $2.15$ & $-1.35$ & $24 \times 12^3$ & $4000$ & $1$ & $2.0$ \\
    DB3M2 & $2.15$ & $-1.36$ & $24 \times 12^3$ & $4000$ & $1$ & $2.0$ \\
    DB3M3 & $2.15$ & $-1.37$ & $24 \times 12^3$ & $4000$ & $1$ & $2.0$ \\
    DB3M4 & $2.15$ & $-1.38$ & $32 \times 16^3$ & $4100$ & $1$ & $2.0$ \\
    DB3M5 & $2.15$ & $-1.39$ & $32 \times 16^3$ & $4077$ & $1$ & $2.0$ \\
    DB3M6 & $2.15$ & $-1.4$ & $32 \times 16^3$ & $4014$ & $1$ & $2.0$ \\
    DB3M7 & $2.15$ & $-1.41$ & $48 \times 24^3$ & $4000$ & $1$ & $2.0$ \\
    DB3M8 & $2.15$ & $-1.422$ & $48 \times 24^3$ & $4000$ & $1$ & $2.0$
\end{tabular}

%% file: assets/tables/lattice_params_Nf1_part1.tex
\begin{tabular}{c|ccc|ccc}
     & $\beta$ & $am$ & $N_t \times N_s^3$ & $N_{\rm conf.}$ & $\delta_{\rm conf.}$ & $\tau_{\rm traj.}$ \\
    \hline
    \hline
    DB4M1 & $2.2$ & $-1.28$ & $24 \times 12^3$ & $8000$ & $1$ & $2.0$ \\
    DB4M2 & $2.2$ & $-1.29$ & $24 \times 12^3$ & $4320$ & $1$ & $2.0$ \\
    DB4M3 & $2.2$ & $-1.3$ & $32 \times 16^3$ & $4000$ & $1$ & $2.0$ \\
    DB4M3* & $2.2$ & $-1.3$ & $24 \times 12^3$ & $4000$ & $1$ & $2.0$ \\
    DB4M4 & $2.2$ & $-1.31$ & $32 \times 16^3$ & $4000$ & $1$ & $2.0$ \\
    DB4M4* & $2.2$ & $-1.31$ & $24 \times 12^3$ & $4000$ & $1$ & $2.0$ \\
    DB4M5 & $2.2$ & $-1.32$ & $32 \times 16^3$ & $4000$ & $1$ & $2.0$ \\
    DB4M5* & $2.2$ & $-1.32$ & $24 \times 12^3$ & $4000$ & $1$ & $2.0$ \\
    DB4M6 & $2.2$ & $-1.33$ & $24 \times 12^3$ & $4092$ & $1$ & $2.0$ \\
    DB4M7 & $2.2$ & $-1.34$ & $32 \times 16^3$ & $4000$ & $1$ & $2.0$ \\
    DB4M7* & $2.2$ & $-1.34$ & $24 \times 12^3$ & $4000$ & $1$ & $2.0$ \\
    DB4M8 & $2.2$ & $-1.35$ & $32 \times 16^3$ & $4000$ & $1$ & $2.0$ \\
    DB4M9 & $2.2$ & $-1.36$ & $48 \times 24^3$ & $4000$ & $1$ & $1.0$ \\
    DB4M9* & $2.2$ & $-1.36$ & $32 \times 16^3$ & $4000$ & $1$ & $2.0$ \\
    DB4M9** & $2.2$ & $-1.36$ & $24 \times 12^3$ & $4000$ & $1$ & $2.0$ \\
    DB4M10 & $2.2$ & $-1.37$ & $48 \times 24^3$ & $4000$ & $1$ & $2.0$ \\
    DB4M11 & $2.2$ & $-1.378$ & $48 \times 24^3$ & $4000$ & $1$ & $1.0$ \\
    DB4M12 & $2.2$ & $-1.386$ & $64 \times 32^3$ & $4226$ & $1$ & $2.0$ \\
    DB4M13 & $2.2$ & $-1.39$ & $96 \times 48^3$ & $4049$ & $1$ & $2.0$ \\
    \hline
    DB5M1 & $2.25$ & $-1.22$ & $24 \times 12^3$ & $4000$ & $1$ & $2.0$ \\
    DB5M2 & $2.25$ & $-1.24$ & $24 \times 12^3$ & $4000$ & $1$ & $2.0$ \\
    DB5M3 & $2.25$ & $-1.26$ & $24 \times 12^3$ & $4000$ & $1$ & $2.0$ \\
    DB5M4 & $2.25$ & $-1.28$ & $32 \times 16^3$ & $4000$ & $1$ & $2.0$ \\
    DB5M4* & $2.25$ & $-1.28$ & $24 \times 12^3$ & $4000$ & $1$ & $2.0$ \\
    DB5M5 & $2.25$ & $-1.3$ & $32 \times 16^3$ & $4000$ & $1$ & $2.0$ \\
    DB5M5* & $2.25$ & $-1.3$ & $24 \times 12^3$ & $4000$ & $1$ & $2.0$ \\
    DB5M6 & $2.25$ & $-1.31$ & $32 \times 16^3$ & $4000$ & $1$ & $2.0$ \\
    DB5M7 & $2.25$ & $-1.32$ & $32 \times 16^3$ & $4000$ & $1$ & $2.0$ \\
    DB5M8 & $2.25$ & $-1.33$ & $32 \times 16^3$ & $4000$ & $1$ & $2.0$ \\
    \hline
    DB6M1 & $2.3$ & $-1.16$ & $32 \times 16^3$ & $4000$ & $1$ & $2.0$ \\
    DB6M1* & $2.3$ & $-1.16$ & $24 \times 12^3$ & $4000$ & $1$ & $2.0$ \\
    DB6M2 & $2.3$ & $-1.18$ & $32 \times 16^3$ & $4000$ & $1$ & $2.0$ \\
    DB6M2* & $2.3$ & $-1.18$ & $24 \times 12^3$ & $4000$ & $1$ & $2.0$ \\
    DB6M3 & $2.3$ & $-1.2$ & $32 \times 16^3$ & $4000$ & $1$ & $2.0$ \\
    DB6M3* & $2.3$ & $-1.2$ & $24 \times 12^3$ & $4000$ & $1$ & $2.0$ \\
    DB6M4 & $2.3$ & $-1.22$ & $32 \times 16^3$ & $4000$ & $1$ & $2.0$ \\
    DB6M5 & $2.3$ & $-1.24$ & $32 \times 16^3$ & $4000$ & $1$ & $2.0$ \\
    DB6M5* & $2.3$ & $-1.24$ & $24 \times 12^3$ & $4000$ & $1$ & $2.0$ \\
    DB6M6 & $2.3$ & $-1.26$ & $32 \times 16^3$ & $4000$ & $1$ & $2.0$ \\
    DB6M6* & $2.3$ & $-1.26$ & $24 \times 12^3$ & $4000$ & $1$ & $2.0$ \\
    DB6M7 & $2.3$ & $-1.28$ & $48 \times 24^3$ & $4000$ & $1$ & $2.0$ \\
    DB6M7* & $2.3$ & $-1.28$ & $32 \times 16^3$ & $4000$ & $1$ & $2.0$ \\
    DB6M8 & $2.3$ & $-1.304$ & $64 \times 32^3$ & $4285$ & $1$ & $2.0$ \\
    DB6M9 & $2.3$ & $-1.31$ & $96 \times 48^3$ & $4141$ & $1$ & $2.0$ \\
    \hline
    DB7M1 & $2.4$ & $-1.1$ & $32 \times 16^3$ & $4000$ & $4$ & $2.0$ \\
    DB7M2 & $2.4$ & $-1.12$ & $32 \times 16^3$ & $4000$ & $4$ & $2.0$ \\
    DB7M3 & $2.4$ & $-1.14$ & $32 \times 16^3$ & $4000$ & $4$ & $2.0$ \\
    DB7M4 & $2.4$ & $-1.16$ & $32 \times 16^3$ & $4000$ & $1$ & $2.0$ \\
    DB7M5 & $2.4$ & $-1.18$ & $32 \times 16^3$ & $4000$ & $1$ & $2.0$ \\
    DB7M6 & $2.4$ & $-1.2$ & $32 \times 16^3$ & $4000$ & $1$ & $2.0$ \\
    DB7M7 & $2.4$ & $-1.21$ & $48 \times 24^3$ & $4090$ & $1$ & $2.0$ \\
    DB7M8 & $2.4$ & $-1.222$ & $48 \times 24^3$ & $4702$ & $1$ & $2.0$ \\
    DB7M9 & $2.4$ & $-1.234$ & $64 \times 32^3$ & $4000$ & $1$ & $1.0$ \\
    DB7M10 & $2.4$ & $-1.243$ & $96 \times 48^3$ & $1736$ & $1$ & $2.0$
\end{tabular}

%% file: assets/plots/polyakov_Nf1_caption.tex
\begin{figure}
  \center
  \includegraphics{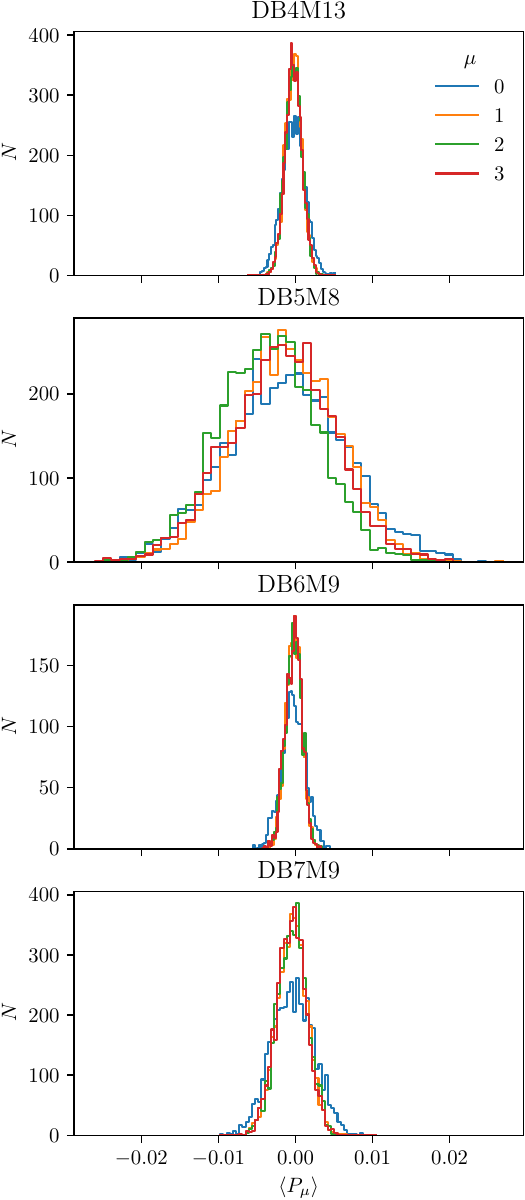}
  \caption{Polyakov loop histograms, for the ensembles DB4M13, DB5M8, DB6M9, and DB7M9.}
  \label{fig:polyakov-Nf1}
\end{figure}

%% file: assets/plots/q_topology_Nf1_caption.tex
\begin{figure}
  \center
  \includegraphics{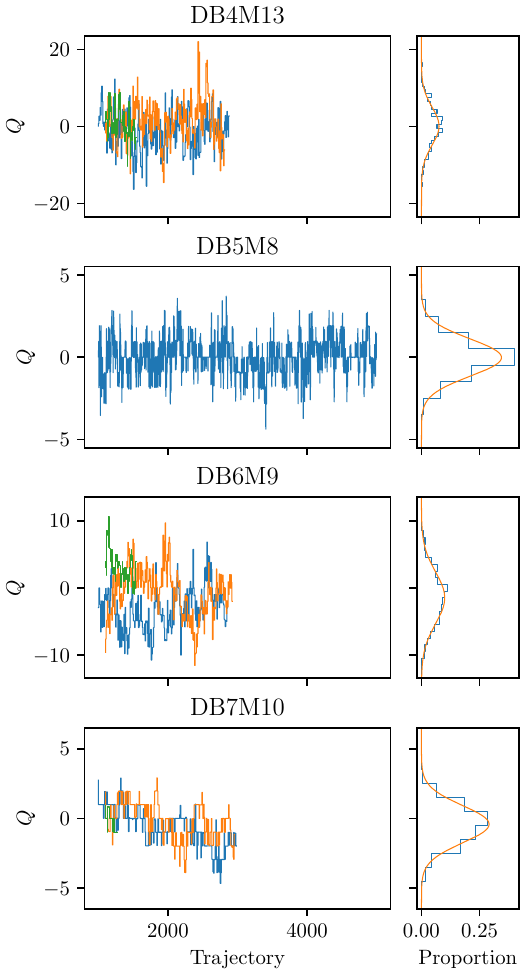}
  \caption{
Topological charge histories (left), and histograms (right), for the ensembles
DB4M13, DB5M8, DB6M9, and DB7M10.}
  \label{fig:topcharge_Nf1}
\end{figure}

%% file: assets/tables/gamma_Nf1.tex
\begin{tabular}{c|cc|cc}
    $\beta$ & $\gamma_*$ (FSHS) & $N_{\mathrm{points}}$ & $\gamma_*$ (AIC) & Ensemble \\
    \hline
    \hline
\multirow{2}{*}{2.05} & \multirow{2}{*}{$0.926(68)$} & \multirow{2}{*}{13} & 0.848(16)(16) & DB1M8 \\
 & & & 0.878(15)(16) & DB1M9 \\
\multirow{1}{*}{2.1} & \multirow{1}{*}{$0.89(12)$} & \multirow{1}{*}{4} & 0.7009(66)(75) & DB2M7 \\
\multirow{1}{*}{2.15} & \multirow{1}{*}{$0.777(47)$} & \multirow{1}{*}{7} & 0.671(36)(21) & DB3M8 \\
\multirow{2}{*}{2.2} & \multirow{2}{*}{$0.693(30)$} & \multirow{2}{*}{23} & 0.589(15)(15) & DB4M11 \\
 & & & 0.5359(16)(13) & DB4M13 \\
2.25 & $0.69(11)$ & 4 & $\cdots$ & $\cdots$ \\
\multirow{1}{*}{2.3} & \multirow{1}{*}{$0.587(19)$} & \multirow{1}{*}{8} & 0.4468(14)(24) & DB6M9 \\
\multirow{2}{*}{2.4} & \multirow{2}{*}{$0.498(19)$} & \multirow{2}{*}{13} & 0.3892(34)(29) & DB7M7 \\
 & & & 0.3843(13)(11) & DB7M10
\end{tabular}

%% file: assets/tables/lattice_params_Nf2_part0.tex
\begin{tabular}{c|ccc|ccc}
     & $\beta$ & $am$ & $N_t \times N_s^3$ & $N_{\rm conf.}$ & $\delta_{\rm conf.}$ & $\tau_{\rm traj.}$ \\
    \hline
    \hline
    Nf2DB0M1 & $2.25$ & $-0.8$ & $48 \times 24^3$ & $4001$ & $2$ & $2.0$ \\
    Nf2DB0M2 & $2.25$ & $-0.9$ & $48 \times 24^3$ & $4001$ & $2$ & $2.0$ \\
    Nf2DB0M3 & $2.25$ & $-0.95$ & $64 \times 32^3$ & $4001$ & $2$ & $2.0$ \\
    Nf2DB0M3* & $2.25$ & $-0.95$ & $48 \times 24^3$ & $4001$ & $2$ & $2.0$ \\
    Nf2DB0M4 & $2.25$ & $-1.0$ & $64 \times 32^3$ & $4001$ & $2$ & $2.0$ \\
    Nf2DB0M4* & $2.25$ & $-1.0$ & $48 \times 24^3$ & $4001$ & $2$ & $2.0$ \\
    Nf2DB0M5 & $2.25$ & $-1.05$ & $96 \times 48^3$ & $4001$ & $1$ & $2.0$ \\
    Nf2DB0M5* & $2.25$ & $-1.05$ & $64 \times 32^3$ & $4001$ & $2$ & $2.0$ \\
    Nf2DB0M6 & $2.25$ & $-1.15$ & $96 \times 48^3$ & $4001$ & $1$ & $2.0$ \\
    \hline
    Nf2DB1M1 & $2.3$ & $-0.2$ & $32 \times 16^3$ & $4001$ & $2$ & $2.0$ \\
    Nf2DB1M2 & $2.3$ & $-0.4$ & $32 \times 16^3$ & $4001$ & $2$ & $2.0$ \\
    Nf2DB1M3 & $2.3$ & $-0.6$ & $32 \times 16^3$ & $4001$ & $2$ & $2.0$ \\
    Nf2DB1M4 & $2.3$ & $-0.8$ & $32 \times 16^3$ & $4001$ & $2$ & $2.0$ \\
    Nf2DB1M5 & $2.3$ & $-0.9$ & $32 \times 16^3$ & $4001$ & $2$ & $2.0$ \\
    Nf2DB1M6 & $2.3$ & $-0.95$ & $32 \times 16^3$ & $4001$ & $2$ & $2.0$ \\
    Nf2DB1M7 & $2.3$ & $-1.0$ & $48 \times 24^3$ & $4001$ & $2$ & $2.0$ \\
    Nf2DB1M7* & $2.3$ & $-1.0$ & $32 \times 16^3$ & $4001$ & $2$ & $2.0$ \\
    Nf2DB1M8 & $2.3$ & $-1.05$ & $48 \times 24^3$ & $4001$ & $2$ & $2.0$ \\
    Nf2DB1M8* & $2.3$ & $-1.05$ & $32 \times 16^3$ & $4001$ & $2$ & $2.0$ \\
    Nf2DB1M9 & $2.3$ & $-1.1$ & $48 \times 24^3$ & $4001$ & $2$ & $2.0$ \\
    Nf2DB1M10 & $2.3$ & $-1.118$ & $48 \times 24^3$ & $4001$ & $2$ & $2.0$ \\
    Nf2DB1M11 & $2.3$ & $-1.12$ & $64 \times 32^3$ & $4001$ & $2$ & $2.0$ \\
    Nf2DB1M12 & $2.3$ & $-1.132$ & $96 \times 48^3$ & $2001$ & $1$ & $2.0$ \\
    Nf2DB1M12* & $2.3$ & $-1.132$ & $64 \times 32^3$ & $4001$ & $2$ & $2.0$ \\
    Nf2DB1M13 & $2.3$ & $-1.14$ & $96 \times 48^3$ & $2001$ & $2$ & $2.0$ \\
    \hline
    Nf2DB2M1 & $2.35$ & $-0.9$ & $32 \times 16^3$ & $4001$ & $1$ & $2.0$ \\
    Nf2DB2M2 & $2.35$ & $-0.95$ & $32 \times 16^3$ & $4001$ & $1$ & $1.0$ \\
    Nf2DB2M3 & $2.35$ & $-1.0$ & $48 \times 24^3$ & $4001$ & $1$ & $1.0$ \\
    Nf2DB2M4 & $2.35$ & $-1.03$ & $48 \times 24^3$ & $4001$ & $1$ & $1.5$ \\
    Nf2DB2M5 & $2.35$ & $-1.05$ & $64 \times 32^3$ & $4086$ & $1$ & $1.0$ \\
    Nf2DB2M6 & $2.35$ & $-1.09$ & $96 \times 48^3$ & $4309$ & $1$ & $2.0$ \\
    Nf2DB2M6* & $2.35$ & $-1.09$ & $64 \times 32^3$ & $4332$ & $1$ & $2.0$ \\
    Nf2DB2M7 & $2.35$ & $-1.11$ & $128 \times 64^3$ & $4428$ & $1$ & $2.0$ \\
    Nf2DB2M7* & $2.35$ & $-1.11$ & $96 \times 48^3$ & $3945$ & $1$ & $2.0$
\end{tabular}

%% file: assets/plots/polyakov_Nf2_caption.tex
\begin{figure}
  \center
  \includegraphics{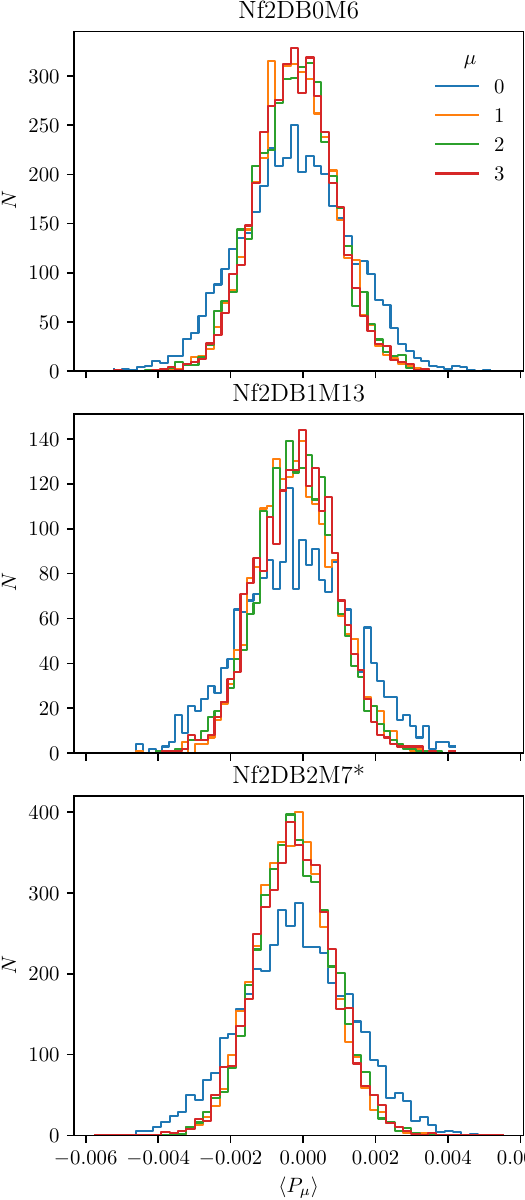}
  \caption{Polyakov loop histograms, for the ensembles Nf2DB0M6, Nf2DB1M13, and Nf2DB2M7*.}
  \label{fig:polyakov-Nf2}
\end{figure}

%% file: assets/plots/q_topology_Nf2_caption.tex
\begin{figure}
  \center
  \includegraphics{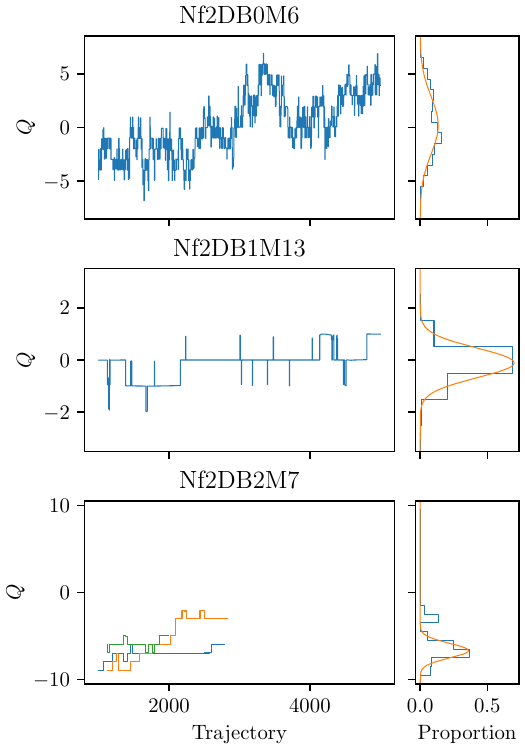}
  \caption{
Topological charge histories (left), and histograms (right), for the ensembles
Nf2DB0M6, Nf2DB1M13, and Nf2DB2M7.}
  \label{fig:topcharge_Nf2}
\end{figure}

%% file: assets/tables/gamma_Nf2.tex
\begin{tabular}{c|cc|cc}
    $\beta$ & $\gamma_*$ (FSHS) & $N_{\mathrm{points}}$ & $\gamma_*$ (AIC) & Ensemble \\
    \hline
    \hline
\multirow{1}{*}{2.25} & \multirow{1}{*}{$0.4593(29)$} & \multirow{1}{*}{11} & 0.3378(16)(46) & Nf2DB0M6 \\
\multirow{1}{*}{2.3} & \multirow{1}{*}{$0.4432(63)$} & \multirow{1}{*}{21} & 0.3093(13)(50) & Nf2DB1M13 \\
\multirow{1}{*}{2.35} & \multirow{1}{*}{$0.3332(61)$} & \multirow{1}{*}{10} & 0.3042(39)(109) & Nf2DB2M6
\end{tabular}